\documentclass[aps,prb,a4paper,twocolumn,nofootinbib,superscriptaddress]{revtex4}

\usepackage[english]{babel}

\usepackage[utf8]{inputenc}

\usepackage{amsmath,amssymb,amsfonts,mathrsfs}

\usepackage[amsmath,thmmarks]{ntheorem}

\usepackage{graphicx}

\usepackage{pdfpages}
\usepackage{wrapfig}
\usepackage{textpos}
\usepackage{subfigure}
\usepackage{float}
\usepackage{hhline}

\usepackage{braket}
\usepackage{upgreek}
\usepackage{lipsum}

\usepackage[]{hyperref}
\usepackage{dsfont}

\newcommand{\vect}[1]{\mathbf{#1}}
\newcommand{\ii}{\mathrm{i}}

\begin{document}

\title{Fluctuation-damping of isolated, oscillating Bose-Einstein condensates}

\author{Tim Lappe}
\email[Email: ]{lappet@th.physik.uni-bonn.de}
\affiliation{Physikalisches Institut and Bethe Center for Theoretical Physics, Universit\"at Bonn, Nussallee 12, 53115 Bonn, Germany} 

\author{Anna Posazhennikova}
\email[Email: ]{anna.posazhennikova@rhul.ac.uk}
\affiliation{Department of Physics, Royal Holloway, University of London, Egham, Surrey TW20 0EX, UK}

\author{Johann Kroha}
\email[Email: ]{kroha@th.physik.uni-bonn.de}

\affiliation{Physikalisches Institut and Bethe Center for Theoretical Physics, Universit\"at Bonn, Nussallee 12, 53115 Bonn, Germany} 

\affiliation{Center for Correlated Matter, Zhejiang University, Hangzhou, Zhejiang 310058, China} 

\date{August 10, 2018} 

\begin{abstract}
Experiments on the nonequilibrium dynamics of an isolated Bose-Einstein
condensate (BEC) in a magnetic double-well trap exhibit a puzzling divergence:
While some show dissipation-free Josephson oscillations, others find strong
damping. Such damping in isolated BECs cannot be understood on the level of
the coherent Gross-Pitaevskii dynamics. Using the Keldysh functional-integral
formalism, we describe the time-dependent system dynamics by means of a
multi-mode BEC coupled to fluctuations (single-particle excitations) beyond
the Gross-Pitaevskii saddle point. We find that the Josephson oscillations
excite an excess of fluctuations when the effective Josephson frequency,
$\tilde{\omega}_J$, is in resonance with the effective fluctuation energy,
$\tilde{\varepsilon}_m$, where both, $\tilde{\omega}_J$ and
$\tilde{\varepsilon}_m$, are strongly renormalized with respect to their
noninteracting values. Evaluating and using the model parameters for the respective experiments describes quantitatively the presence or absence of damping.
\end{abstract}

\maketitle


\section{Introduction}\label{sec:intro}

When a system of ultracold, condensed bosons is trapped in a double-well
potential with an initial population imbalance, it undergoes Josephson
oscillations\cite{Smerzi1997} between the wells and can, therefore, be 
referred to as a Bose-Josephson junction (BJJ). 
Josephson oscillations were observed in a number of
experiments.\cite{Albiez2005,Levy2007,LeBlanc2011,Spagnolli2017} 
Since the experimental
systems are almost ideally separated from the environment, a BJJ can serve as a
prototype of a nonequilibrium closed quantum system. Because of the
unitary time-evolution which prohibits the maximization of entropy,
a closed quantum system cannot thermalize as a whole, once driven out 
of equilibrium. However, strong damping of Josephson oscillations was observed 
in the experiments by LeBlanc {\it et al.},\cite{LeBlanc2011} whereas the 
experiments by Albiez {\it et al.}\cite{Albiez2005} and by 
Spagnolli {\it et al.} clearly displayed undamped 
oscillations for extended periods of time. 
Explaining this discrepancy and, thereby, giving guidelines for  
designing an experimental setup with or without damping and thermalization, 
is the aim of this work.  

Previously some of the present authors proposed the dynamical heat-bath
generation (DBG) as a damping and thermalization
mechanism:\cite{Posazhennikova2018, Posazhennikova2016} For a sufficiently
complex, isolated quantum system the Hilbert space dimension is so large
that only a small subset of the huge amount
of quantum numbers characterizing the system's state vector can be 
determined in any given experiment. This subset defines a subspace of the 
total Hilbert space, referred to as the ``subsystem'' $\mathcal{S}$. 
Any measurement performed on ${\cal S}$ alone is partially destructive,
in that the quantum numbers defining the Hilbert space of ${\cal S}$
are fixed (partial state collapse), but the remaining subspace of
undetermined quantum numbers is traced out. This remaining subspace,
${\cal R}$, becomes massively entangled\cite{Lenk2018} with the states of the
subsystem ${\cal S}$ via the many-body dynamics and, hence, acts as a
grand-canonical bath or reservoir.
By the resulting, effectively grand canonical time evolution of 
the subsystem $\mathcal{S}$, it will naturally reach a thermal state 
in the long-time limit,\cite{Posazhennikova2016} if the system is ergodic. 
Thus, the measurement process itself defines a division into subsystem 
and reservoir. For instance, when the population imbalance in a BJJ 
is measured, the Bose-Einstein condensate (BEC) states comprise $\mathcal{S}$, 
and all the many-body states involving incoherent excitations outside 
the BEC comprise $\mathcal{R}$. 
Note that this thermalization process is dynamical and is possible 
even when the bath states (for a BJJ, the incoherent excitations) are 
initially not occupied, hence the term dynamical bath generation. 
By contrast, the so-called eigenstate thermalization 
hypothesis\cite{Deutsch1991,Srednicki1994} (ETH) requires the 
system to be near a many-body eigenstate of the total Hamiltonian
(microcanonical ensemble), i.e., it is stationary by construction. See 
Ref.~[\onlinecite{Posazhennikova2018}] for a detailed discussion.

The DBG mechanism was corroborated for a BJJ with arbitrary system 
parameters, where it was shown that incoherent excitations  
are efficiently generated out of the oscillating BEC due to a parametric 
resonance.\cite{Posazhennikova2016} 
The complex thermalization dynamics of a BJJ involving several time scales
has been analyzed in detail in 
Refs.~[\onlinecite{Posazhennikova2016,Posazhennikova2018}]. In particular, 
the thermalization time $\tau_{th}$ is necessarily much larger than the 
BJJ oscillation period, because (1) the incoherent fluctuations are created by 
the Josephson oscillations themselves and (2) because of the quasi-hydrodynamic 
long-time dynamics.\cite{Posazhennikova2016}

In the present work we examine this damping mechanism for realistic 
experimental parameters and specific traps. Previous studies within 
the two-mode approximation\cite{Smerzi1997,Ananikian2006,LeBlanc2011} 
showed significant, interaction-induced renormalizations of the Josephson 
frequency, $\tilde\omega_J$, but did not explain the observed oscillation
damping.\cite{LeBlanc2011}   
A multimode expansion of the Gross-Pitaevskii equation (GPE) 
in terms of the complete basis of single-particle trap eigenmodes
can describe the coherent part of the dynamics in principle exactly.
However, the dynamical excitation of higher trap levels also involves
the creation of {\it incoherent} fluctuations which are not captured
by the GPE saddlepoint dynamics. The excitation
of higher trap modes and the concatenated creation of incoherent
fluctuations is crucial for damping in realistic systems. These
fluctuations are captured by the systematic expansion about the
GPE saddlepoint (see Sec.~\ref{sec:formalism} B), involving BEC as well 
as fluctuation Green's functions.
We find that efficient coupling to higher trap modes occurs 
if $\tilde\omega_J$ is in resonance with the excitation energy of 
one of the trap levels, $\tilde\omega_J\approx\tilde\varepsilon_m$, where 
$\tilde\omega_J$, as well as $\tilde\varepsilon_{m}$, are strongly 
renormalized and broadened by their mutual coupling and by the interactions.
Conversely, in the off-resonant regime, the Josephson oscillations remain 
undamped over an extended period of time. 
Our quantitative calculations reveal that the experimental parameters 
of LeBlanc {\it et al.}\cite{LeBlanc2011} are in the strongly damped regime 
and those of Albiez {\it et al.}\cite{Albiez2005} in the undamped regime, 
in agreement with the experimental findings. 
This reconciles the apparent discrepancy between these two classes of 
experiments and supports the validity of the DBG mechanism in 
Bose-Josephson junctions.

The article is organized as follows. In Sec. \ref{sec:formalism} we describe
the many-body action used to model the system and its representation in the 
trap eigenbasis. We develop the nonequilibrium temporal dynamics by means
of the Keldysh path integral. Sec. \ref{sec:results} contains the numerical
analysis: the resonance effect responsible for the damping, 
and a detailed application to the two exemplary
experiments, Refs.~[\onlinecite{Albiez2005}] and [\onlinecite{LeBlanc2011}],
respectively. This is followed by a discussion and concluding remarks in 
Sec.~\ref{sec:discussion}.


\section{Formalism}\label{sec:formalism}

The model for an ultracold gas in a double-well 
trap potential $V_{\text{ext}}(\vect{r})$ with multiple single-particle 
levels is defined using the functional-integral 
formalism. It allows for a convenient distinction between the 
condensate amplitudes in each level, defined by the time-dependent 
Gross-Pitaevskii saddle point, and the non-condensate excitations. The
nonequilibrium dynamics will be described by the functional integral on the Keldysh time contour.

\subsection{Multi-mode model}
\label{subsec:model}

The action $S$ for a trapped, atomic Bose gas with a contact interaction reads in terms of the bosonic fields $\psi(\vect{r}, t),\psi^*(\vect{r},t)$, 
\begin{align}\label{eq:action}
S\left[\psi,\psi^* \right]&=\int \text{d}^3 r\text{d}t\,\big[ \psi^*(\vect{r},t)G_0^{-1}(\vect{r},t)\psi(\vect{r},t)\nonumber\\
&-\frac{\tilde{g}}{2}\psi^*(\vect{r},t)\psi^*(\vect{r},t)\psi(\vect{r},t)\psi(\vect{r},t)\big],
\end{align}
where the coupling parameter $\tilde{g} = 4\pi \hbar^2 a_s/m$ is proportional to the $s$-wave scattering length $a_s$,\cite{Vogels1997,VanKempen2002} and the inverse free Green function is
\begin{align}\label{eq:invGreen}
G_0^{-1}(\vect{r}, t) =\ii\partial_t - \left(-\frac{\hbar^2\nabla^2}{2m} + V_{\text{ext}}(\vect{r})\right).
\end{align}
The spatial dependence of the field $\psi(\vect{r},t)$ may be resolved into the complete, orthonormal basis of single-particle 
eigenfunctions $\{\varphi_-(\vect{r}),\varphi_+(\vect{r}),\varphi_3(\vect{r}),\varphi_4(\vect{r}),\,\dots\}$ of the trap,\cite{Posazhennikova2016}  
\begin{align}\label{eq:spectralSum}
	\begin{split}
		\psi(\vect{r},t) &= \psi_+(\vect{r},t) + \psi_-(\vect{r},t) + \sum_{m=3}^{M} \psi_m(\vect{r},t) \\ 
        &= \varphi_+(\vect{r})\phi_+(t) + \varphi_-(\vect{r})\phi_-(t) + \sum_{m=3}^{M} \varphi_m(\vect{r})\phi_m(t),
	\end{split}
\end{align}
with time-dependent amplitudes $\phi_m(t)$ and $M$ the number of modes 
taken into account. The $\varphi_i(\vect{r})$ are the solutions of the 
stationary Schr\"odinger equation with the potential $V_{\text{ext}}(\vect{r})$,
with eigenfrequencies 
$\{\varepsilon_-,\,\varepsilon_+,\,\varepsilon_3,\,\varepsilon_4, \dots\}$. The wavefunctions
$\varphi_-(\vect{r})$ and $\varphi_+(\vect{r})$ are the two lowest-lying 
eigenfunctions of $V_{\text{ext}}(\vect{r})$ extending over both wells, 
with odd (-) and even (+) parity, respectively.  
In view of the anticipated dynamics with different occupation numbers 
in the two wells, it is useful to define the symmetric and antisymmetric 
superpositions 
$\varphi_{1,2}(\vect{r})=[\varphi_-(\vect{r}) \pm \varphi_+(\vect{r})]/\sqrt{2}$,
since they are localized in the left or right well, respectively.  
With the expansion (\ref{eq:spectralSum}) the action takes the form $S=S_0+S_{\mathrm{int}}$, with the noninteracting part,
\begin{align}
S_{0}=\int \text{d}t\left\{\sum_{i=1}^{M}\left[\phi_{i}^{*}\left(\ii\partial_t -\varepsilon_i\right)\phi_{i}^{}\right] - J\left( \phi_{1}^{*}\phi_{2}^{} + \phi_{2}^{*}\phi_{1}^{} \right)\right\}, 
\label{eq:S0}
\end{align}
and the interacting part
\begin{align}
S_{\text{int}} = -\tfrac{1}{2}\int \text{d}t \sum_{ijkl=1}^{M}U_{ijkl} \, \phi_i^{*}(t) \phi_j^{*}(t) \phi_k^{}(t) \phi_l^{}(t),
\label{eq:Sint}
\end{align}
where the $\phi_{1,\,2}^{}$ are the symmetric and antisymmetric superpositions 
of the time-dependent fields $\phi_{\pm}^{}(t)$.
In this mode representation, 
the spatial dependence of the Bose field $\psi(\vect{r},t)$ 
is absorbed into the overlap integrals $\varepsilon_i$, $J$, and $U_{ijkl}$,
which are given by
\begin{align}
\varepsilon_{i}=&\int \text{d}^3 r\,\varphi_i^*(\vect{r})\,
\left(-\frac{\hbar^2\nabla^2}{2m} + V_{\text{ext}}(\vect{r})\right)\,
\varphi_i(\vect{r}),\ \ \ \label{eq:eps}\\
\varepsilon_1 =& \,\varepsilon_2=\tfrac{1}{2}(\varepsilon_{-}+\varepsilon_{+}),\\
J=&\int \text{d}^3 r\,\varphi_1^*(\vect{r})\,
\left(-\frac{\hbar^2\nabla^2}{2m} + V_{\text{ext}}(\vect{r})\right)\,
\varphi_2(\vect{r})\label{eq:J}\nonumber\\
=&\tfrac{1}{2}(\varepsilon_{-}-\varepsilon_{+}),\\
U_{ijkl}=&\tilde{g}\int \text{d}^3 r\,
\varphi_i^*(\vect{r})\varphi_j^*(\vect{r})\varphi_k(\vect{r})\varphi_l(\vect{r}),
\label{eq:U}	
\end{align}
where the bound-state functions $\varphi_i(\vect{r})$ may be chosen real.
Note that a bare Josephson coupling $J$ exists only between the two lowest 
modes $\varphi_1(\vect{r})$, $\varphi_2(\vect{r})$, localized in the 
left or right well, while the modes with $i\geq 3$ are trap eigenmodes 
and extended over the entire trap. Without loss of generality we may choose 
the zero of energy as $\varepsilon_1=\varepsilon_2=0$.

For $M\to\infty$ the representation Eqs.~(\ref{eq:S0})--(\ref{eq:U}) 
in terms of the single-particle trap eigenmodes is exact.
Numerically, the decomposition in Eq. \eqref{eq:spectralSum} is analogous to a Galerkin method. Replacing the space-dependence by summations over eigenfunctions leads to a significant simplification of the numerical initial-value problem when truncating the decomposition at a finite value of $M$. In this work we will take $M=4,\, 6$, depending on the form of the external potential $V_{\text{ext}}(\vect{r})$, see section \ref{sec:results}.

\subsection{Nonequilibrium effective action}
\label{sec:keldyshformalism}

In this subsection, we are going to present the formal derivation of the equations of motion in the Bogoliubov-Hartree-Fock (BHF) approximation that describe the condensate and its exchange with a cloud of noncondensed particles.

The Keldysh technique\cite{Keldysh1964} in path-integral formulation\cite{Sieberer2016} is a particularly elegant tool for the construction of self-consistent approximations via the effective action, where both the condensate amplitudes $\Phi_i = \langle \phi_i\rangle$ and the fluctuations above the condensate, $\delta\phi_i$, are treated on an equal footing. 

For the general derivation of the one-particle irreducible (1PI) effective action, we will suppress the field indices and instead work with a time-dependent field $\phi$ which can in principle carry arbitrary quantum numbers. The bosonic fields should now be separated into fields on the forward branch $C_1$ of the Keldysh contour and fields on the backward branch $C_2$, such that we can express the action as
\begin{align}
S_{K}[\phi^{}_{C_1},\phi^{*}_{C_1},\phi^{}_{C_2},\phi^{*}_{C_2}]=S[\phi^{}_{C_1},\phi^{*}_{C_1}] - S[\phi^{}_{C_2},\phi^{*}_{C_2}].
\end{align}
From this action we obtain $S_{K}[\phi^{}_{c},\phi^{*}_{c},\phi^{}_{q},\phi^{*}_{q}]$ by performing the Keldysh rotation according to
\begin{align}
\phi^{}_{C_1}=\tfrac{1}{\sqrt{2}}(\phi^{}_{c}+\phi^{}_{q}), \quad \phi^{}_{C_2}=\tfrac{1}{\sqrt{2}}(\phi^{}_{c}-\phi^{}_{q}),
\end{align}
where $c$ stands for "classical" and $q$ for "quantum". This nomenclature stems from the fact that neglecting fluctuations, the field $\phi^{}_c$ will obey the classical equations of motion which follow from the corresponding classical action. The ``quantum'' field $\phi^{}_q$ is the so-called ``response'' field describing all fluctuations (both classical and quantum). In the simplest classical limit, it essentially corresponds to a description of Gaussian white noise with zero mean through the characteristic functional known from probability theory. 

Defining complex field spinors $\boldsymbol{\phi}=\left( \phi, \phi^{*}_{} \right)^{T}$ and external sources $\boldsymbol{j}=\left( j, j^{*}_{} \right)^{T}$, the partition function will be 
\begin{align}\label{eq:partFunc}
Z[\boldsymbol{j}^{}_{c},\boldsymbol{j}^{}_{q}]&=\int \mathcal{D}[\boldsymbol{\phi}^{}_{c},\boldsymbol{\phi}^{}_{q}]e^{\ii S_{K}[\boldsymbol{\phi}^{}_{c},\boldsymbol{\phi}^{}_{q}]} e^{\ii\hspace*{-0.05cm}\int \text{d}t(\boldsymbol{j}^{\dagger}_{q}\boldsymbol{\phi}^{}_{c} + \boldsymbol{j}^{\dagger}_{c}\boldsymbol{\phi}^{}_{q})},
\end{align}
where we have also introduced Keldysh classical and quantum components for the external sources. Taking the logarithm of $Z$, we find the cumulant-generating functional
\begin{align}
W[\boldsymbol{j}^{}_{c},\boldsymbol{j}^{}_{q}]=-\ii\ln{Z[\boldsymbol{j}^{}_{c},\boldsymbol{j}^{}_{q}]}.
\end{align}
Differentiation with respect to $\boldsymbol{j}$ gives the expectation value of the field in the presence of external sources,
\begin{align}
\Phi_{c,\,q}=\langle\phi_{c,\,q}\rangle=\frac{\delta W}{\delta j_{q,\,c}^{*}}.
\end{align}
and we define $\boldsymbol{\Phi} = \left( \Phi, \Phi^{*}_{} \right)^{T}$. By a Legendre transform\cite{Jackiw1974} to these new variables, we arrive at the 1PI effective action 
\begin{align}\label{eq:effAction}
\Gamma[\boldsymbol{\Phi}^{}_{c},\boldsymbol{\Phi}^{}_{q}]=W[\boldsymbol{j}^{}_{c},\boldsymbol{j}^{}_{q}]-\int \mathrm{d}t(\boldsymbol{j}^{\dagger}_{q}\boldsymbol{\Phi}^{}_{c} + \boldsymbol{j}^{\dagger}_{c}\boldsymbol{\Phi}^{}_{q}),
\end{align}
which will be the main tool of our analysis, since it  allows for a rigorous derivation of self-consistent perturbation theory. To this end, we finally decompose the field into a finite average plus fluctuations according to
\begin{align}\label{eq:flucs_decomp}
\phi_{c,\,q}=\Phi_{c,\,q}+\delta\phi_{c,\,q}.
\end{align}
Plugging this into Eq. \eqref{eq:effAction}, and using \eqref{eq:partFunc}, the source terms coupled to the averages $\boldsymbol{\Phi}$ vanish, and we are left with
{
\begin{align}
\begin{split}
& e^{\ii\Gamma[\boldsymbol{\Phi}^{}_{c},\boldsymbol{\Phi}^{}_{q}]}=\int \mathcal{D}[\delta\boldsymbol{\phi}^{}_{c}, \delta\boldsymbol{\phi}^{}_{q}]  \exp{\left\{\ii S_{K}[\boldsymbol{\phi}^{}_{c} ,\boldsymbol{\phi}^{}_{q}]\right\}}\\
&\times \exp{\left\{-\ii\int\text{d}t\left(\left(\frac{\delta\Gamma}{\delta\boldsymbol{\Phi}^{}_{c}}\right)^{T}\delta\boldsymbol{\phi}^{}_{c} 
+\left(\frac{\delta\Gamma}{\delta\boldsymbol{\Phi}^{}_{q}}\right)^{T}\delta\boldsymbol{\phi}^{}_{q}\right)\right\}}.
\end{split}
\end{align}
}
This path integral supplements the field averages by fluctuation terms in a similar way to a Ginzburg-Landau approach. 
From it, one can easily generate the established Bogoliubov-Hartree-Fock (BHF) approximation by keeping only the quadratic fluctuations. 
In Appendix A, its conserving properties, that is, conservation of energy and particle number, are proved explicitly for the two-mode model. As is well-known, the BHF approximation is not gapless and violates the Hugenholtz-Pines theorem.  

Variation of the effective action with respect to $\Phi^{*}_{i\,q}$ results in a modified Gross-Pitaevskii equation (GPE), which gives the evolution of the classical average $\Phi_{i\,c}(t)$, describing the condensate,
\begin{align}\label{eq:SPAc}
0&=\frac{\delta\Gamma}{\delta\Phi_{i\,q}^{*}},
\end{align}
while the average of the quantum component has to vanish identically, 
\begin{align}
	\Phi_{i\,q}(t)=0.
\end{align}

Since we would like to investigate the occupation dynamics including the fluctuations, we have to consider the Keldysh Green functions as well, which we write as
{\small
\begin{align}
\vect{G}^{K}_{ij}(t,t')&=
\begin{pmatrix}
    G_{ij}(t,t')  & g_{ij}(t,t') \\
    -g^{*}_{ij}(t,t')  & -G^{*}_{ij}(t,t') \\   
\end{pmatrix}\nonumber\\
&=-\ii\begin{pmatrix}
    \langle\delta\phi_{i\,c}^{}(t)\delta\phi_{j\,c}^{*}(t')\rangle  &  \langle \delta\phi_{i\,c}^{}(t)\delta\phi_{j\,c}^{}(t')\rangle  \\
    \langle \delta\phi_{i\,c}^{*}(t)\delta\phi_{j\,c}^{*}(t')\rangle   &  \langle\delta\phi_{i\,c}^{*}(t)\delta\phi_{j\,c}^{}(t')\rangle    
\end{pmatrix},
\end{align}
}%
where for the matrix elements we drop the Keldysh superscript and explicitly keep the anomalous contributions, designated by a lowercase $g$. Since $\Phi_{i\,q}(t)=0$, in the following we will simply write $\Phi_{i\,c}(t)=\Phi_{i}$. With these definitions, in its most general form Eq. \eqref{eq:SPAc} will be given by
\begin{align}\label{eq:SPA_general}
\begin{split}
0 & =\left( \ii \delta_{ij}\partial_t - h_{ij}\right)\Phi_{j}^{} - \frac{U_{ijkl}}{2}\left[\right. \Phi_{j}^{*}\Phi_{k}^{}\Phi_{l}^{} \\
&+ \ii \Phi_{j}^{*}g_{kl}^{}(t,t) + \ii \Phi_{k}^{}G_{jl}^{}(t,t) + \ii \Phi_{l}^{}G_{jk}^{}(t,t) \left.\right],
\end{split}
\end{align}
where $h_{ij}$ represents the coefficients from the quadratic part of the action, and repeated indices are summed over. Without the contributions from the fluctuations, this would be the standard GPE.

In order to determine the fluctuation Green functions, we have to solve the respective Dyson equations,
{\small
\begin{align}\label{eq:dyson}
\begin{split}
\int \text{d}\bar{t}\;\delta(t-\bar{t})\left([\vect{G}^{R}_{0}]_{ij}^{-1}(t) - \vect{\Sigma}_{ij}^{R}(t) \right)\vect{G}^{K}_{jk}(\bar{t},t')  &=0, \\
\int \text{d}\bar{t}\;\delta(\bar{t}-t') \vect{G}^{K}_{ij}(t, \bar{t})\left([\vect{G}^{A}_{0}]_{jk}^{-1}(t') - \vect{\Sigma}_{jk}^{A}(t')\right) &=0,
\end{split}
\end{align}
}%
self-consistently alongside Eq. \eqref{eq:SPA_general}. The inverse Green functions and self-energies can be obtained from the second derivatives of the effective action,
{\small
\begin{align*}\label{eq:G_inv}
[\vect{G}^{R}_{0}]_{ij}^{-1}(t, t') - \vect{\Sigma}_{ij}^{R}(t, t')&=\begin{pmatrix}
    \frac{\delta^{2}\Gamma}{\delta\Phi_{i\,q}^{*}(t)\delta\Phi_{j\,c}^{}(t')}  &  \frac{\delta^{2}\Gamma}{\delta\Phi_{i\,q}^{*}(t)\delta\Phi_{j\,c}^{*}(t')} \\ \frac{\delta^{2}\Gamma}{\delta\Phi_{i\,q}^{}(t)\delta\Phi_{j\,c}^{}(t')}
   & \frac{\delta^{2}\Gamma}{\delta\Phi_{i\,q}^{}(t)\delta\Phi_{j\,c}^{*}(t')} \\    
    \end{pmatrix}.
\end{align*}
}%
At  Hartree-Fock level, the self-energies are local in time, which leads to the temporal delta functions in \eqref{eq:dyson}. The inverse Green functions are
{\small
\begin{align}\label{eq:inv_green}
[\vect{G}^{R}_{0}]_{ij}^{-1}(t) 
    &= \begin{pmatrix}
    \ii\delta_{ij}\partial_t - h_{ij} &   0\\ 0
   & -\ii\delta_{ij}\partial_t - h_{ij} \\
    \end{pmatrix}, \\
    [\vect{G}^{A}_{0}]_{ij}^{-1}(t) &=
    \begin{pmatrix}
    -\ii\delta_{ij}\overleftarrow{\partial}_{t} - h_{ij} &  0 \\0
   & \ii\delta_{ij}\overleftarrow{\partial}_{t} - h_{ij}  \\
    \end{pmatrix} ,
\end{align}
}%
and the retarded and advanced self-energies read
{\small
\begin{align}\label{eq:selfEn}
\vect{\Sigma}_{ij}^{R}(t) 
    &=  \vect{\Sigma}_{ij}^{A}(t)=\begin{pmatrix}  
     \Sigma_{ij}(t) &   \sigma_{ij}^{}(t)  \\ \sigma_{ij}^{*}(t) 
   & \Sigma_{ij}(t)  \\
    \end{pmatrix} ,
\end{align}
}%
where
{
\begin{align}
\Sigma_{ij}(t)  &=U_{ijkl}\left[ \Phi_k^*(t)\Phi_l^{}(t) + \ii G_{kl}(t,t) \right] ,\\
\sigma_{ij}(t)  &= \frac{U_{ijkl}}{2}\left[ \Phi_k^{}(t)\Phi_l^{}(t) + \ii g_{kl}(t,t) \right].
\end{align}

This set of self-consistent BHF equations for the field averages and the Keldysh components of the Green functions, Eqs. \eqref{eq:SPA_general} and \eqref{eq:dyson}, can be solved in the equal-time limit by combining the retarded and advanced equations.\cite{Trujillo-Martinez2015} Specifically, the upper left and right components of the retarded Bogoliubov-matrix equation in \eqref{eq:dyson} are
\begin{align}\label{eq:retared_general}
\begin{split}
0 &= \left(\ii \delta_{ij}\partial_t - h_{ij} - \Sigma_{ij}(t)\right)G_{jk}(t,t') + \sigma_{ij}(t)g^{*}_{jk}(t,t'), \\
0 &= \left(\ii \delta_{ij}\partial_t - h_{ij} - \Sigma_{ij}(t)\right)g_{jk}(t,t') + \sigma_{ij}(t)G^{*}_{jk}(t,t'),
\end{split}
\end{align}
respectively. Accordingly, the upper left and right components of the advanced equation in \eqref{eq:dyson} are
\begin{align}\label{eq:advanced_general}
\begin{split}
0 &= \left(-\ii \delta_{jk}\partial_{t'} - h_{jk} - \Sigma_{jk}(t')\right)G_{ij}(t,t') - \sigma_{jk}^{*}(t')g_{ij}(t,t'), \\
0 &= \left(\ii \delta_{jk}\partial_{t'} - h_{jk} - \Sigma_{jk}(t')\right)g_{ij}(t,t') - \sigma_{jk}(t)G_{ij}(t,t').
\end{split}
\end{align}
}%
Note the differing time derivatives and arguments of the self-energies. By subtracting the first of Eqs. \eqref{eq:advanced_general} from the first of Eqs. \eqref{eq:retared_general} and taking the equal-time limit, one finds equations for the $G_{ij}(t,t)$. Similarly, by adding the second of Eqs. \eqref{eq:retared_general} to the second of Eqs. \eqref{eq:advanced_general}, in the equal-time limit one obtains equations for the anomalous Green functions $g_{ij}(t,t)$.

Further details of the derivation are exemplified in Appendix A for the two-mode case.

\begin{figure}[t]
\centering    
\subfigure{\label{fig:levels_Obi}\includegraphics[width=0.48\textwidth]{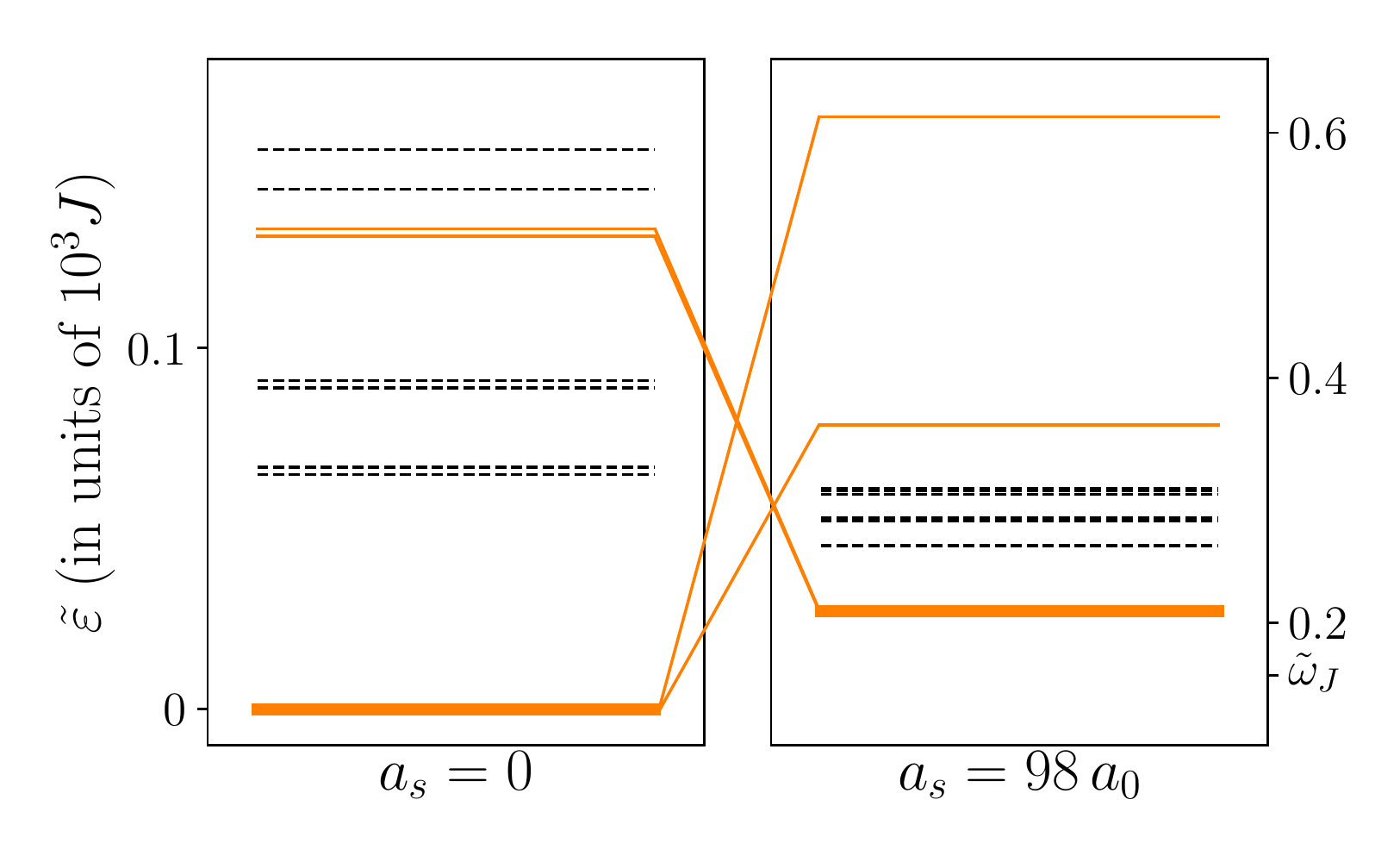}}
\subfigure{\label{fig:levels_Thy}\includegraphics[width=0.48\textwidth]{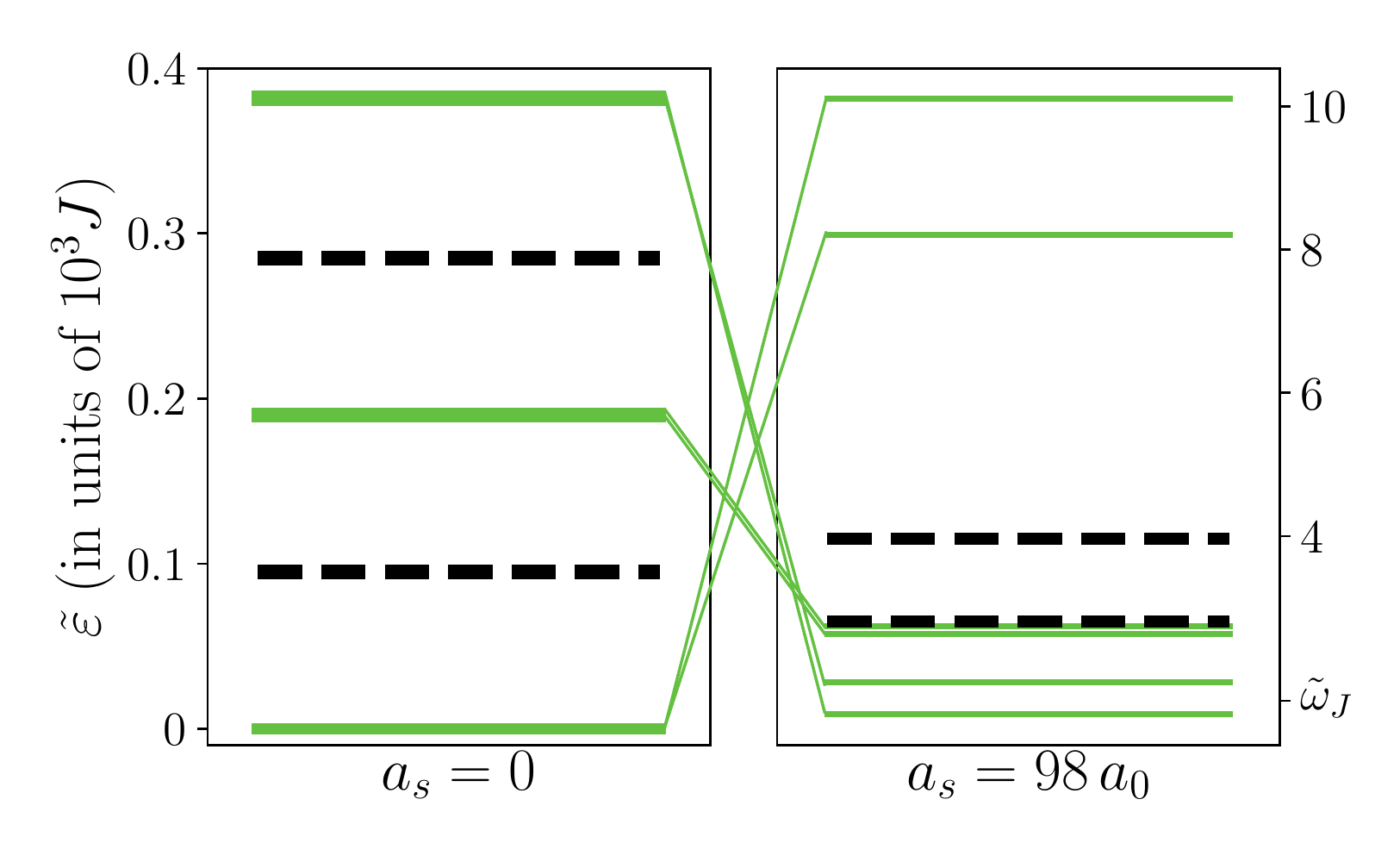}}
\vspace{-0.7cm}
\caption{Bare (left) vs. mean-field-shifted (right) single-particle energies
  of the trapping potentials $V_A({\bf r})$ from
  Ref.~[\onlinecite{Albiez2005}] (upper panels) and $V_B({\bf r})$ from
  Ref.~[\onlinecite{LeBlanc2011}] (lower panels). The first ten levels are shown. Thick lines indicate nearly degenerate state pairs. The right panels show the initial renormalization of the levels due to the interaction ($a_s=98\, a_0$, with $a_0$ the Bohr radius). The solid (colored) lines in the right panels are the ones used for the time-dependent numerical calculations (see text). The renormalized Josephson frequency $\tilde\omega_J$, as extracted from the time evolution of $z(t)$, is also shown for each case.}\label{fig:levels}
\end{figure}


\section{Application to experiments}\label{sec:results}

This section is divided into three parts. The first part is dedicated to the
quantitative calculation of the trap and interaction parameters for the
experiments of Albiez {\it et al.}\cite{Albiez2005} and  
LeBlanc {\it et al.}\cite{LeBlanc2011}, respectively. 
In the second part, by scanning through realistic trap-parameter values, we demonstrate numerically that efficient damping 
can occur only if the resonance condition for the Josephson frequency
$\tilde{\omega}_J$ and the broadened energy levels of the incoherent excitations
$\tilde{\varepsilon}_m$ is fulfilled. 
The third and final part contains our numerical results for experiments 
with  undamped\cite{Albiez2005} and strongly damped\cite{LeBlanc2011} 
Josephson oscillations, respectively.

\subsection{Realistic trap parameters and level renormalization}
\label{subsec:parameters}

We quantitatively analyze two classes experiments: those of Albiez \textit{et al.}\cite{Albiez2005} as an exemplary observation of undamped Josephson oscillations, hereafter referred to as experiment (A), and those by 
LeBlanc \textit{et al.}\cite{LeBlanc2011} where strong damping occurred, 
and which we will refer to as experiment (B). 
Both experiments were performed in double-well potentials, and the population
imbalance $z(t)$ between the two wells was traced as a function of time. 
While the experiments (A) are well described by an effective nonpolynomial 
Schr\"odinger equation,\cite{Salasnich2002}
in the experiments (B) the Fourier transform of $z(t)$ exhibits two or three 
frequencies in addition to damping,\cite{LeBlanc2011} indicating  
contributions from more than two modes.

In order to reduce the numerical effort for the subsequent, time-dependent 
computations, one should select those levels which participate 
significantly in the dynamics. To this end, it is important to realize 
that both, 

\begin{table}[h]
\vspace*{0.3cm}
\centering
\begin{tabular}{|c|c|c|c|c|c|}
\hline
  & $J$    & $U$  & $J'$  & $U'$ & $N$ \\ 
\hhline{|=|=|=|=|=|=|}
Albiez \textit{et al.}\cite{Albiez2005}  &\ $-1.0$ \  &\ $0.40$ \  &\ $-0.002$ \ &\ $0.0001$\  &\  $1150$\  \\ \hline
LeBlanc \textit{et al.}\cite{LeBlanc2011}&\ $-1.0$ \  &\ $1.73$ \  &\ $-0.006$ \ &\ $0.0001$\  &\  $4500$\  \\ \hline
\end{tabular}
\caption{Hamiltonian matrix elements involving the only left- and right-localized modes, and total particle number $N$ for the experiments (A) 
of Albiez \textit{et al.} and (B) of LeBlanc \textit{et al.}, respectively.}
\label{tab:two-mode_params}
\end{table}
\noindent  
the single-particle level energies and the Josephson frequency, are strongly 
renormalized by the interactions. We calculate the level renormalizations 
within the BHF approximation at the initial time $t=0$.
The bare ($\varepsilon$) and the renormalized ($\tilde\varepsilon$) 
single-particle levels 
are shown in Fig.~\ref{fig:levels} for the experiments (A) and (B),
respectively, for the example that all particles are 
initially condensed in the left potential well. It is seen that the 
interactions even change the sequence of the trap levels. In particular, 
the two low-lying left- or right-localized levels ($\alpha =1,2$) 
are shifted upward above the other renormalized levels. 
The reason is that the two lowest-lying single-particle orbitals are 
macroscopically occupied by the BEC atoms with condensate population number
$N_{\alpha}$, so that the energy for adding one additional particle in these 
levels is renormalized on the order of 
$\tilde\varepsilon_{\alpha} \approx \varepsilon_{\alpha} + N_{\alpha} U$, 
with additional contributions from the inter-level condensate 
interactions $U', J'$. 
Similarly, the excited single-particle levels are renormalized predominantly 
by their interaction with the condensates as 
$\tilde\varepsilon_{n} \approx \varepsilon_{n} + N K_n$, $n=3,\,4,\, 5,\dots$, 
where $N=N_1+N_2$ is the total condensate occupation number and $K_n$ substantially smaller than $U$. The interaction-induced re-ordering of levels shown in 
Fig.~\ref{fig:levels} remains valid as long as the ground-state occupations 
$N_{\alpha}$, $\alpha=1,\,2$, are substantial.
As a side result, this level re-ordering justifies the frequently used 
Bogoliubov 
approximation,\cite{Trujillo-Martinez2009,Posazhennikova2016,Posazhennikova2018}
where non-condensate amplitudes
in the left- or right-localized ground modes $\alpha=1,\,2$ are 
neglected, because such fluctuations are energetically suppressed.   
Our calculations show that different initial BEC population 
imbalances $z(0)$ do not significantly alter the 
renormalized level schemes. 
In particular, we find that this remains true 
for the time evolution in both experiments, (A) and (B).   
Therefore, the initial level renormalization shown in Fig.~\ref{fig:levels} 
may be used for selecting the relevant levels at all times during the 
evolution, see subsection \ref{subsec:comparison}.

\onecolumngrid

\begin{table}[b!]
\centering
\begin{tabular}{|c|c|c|c|c|c|c|c|c|c|c|c|c|c|}
\hline
& $\varepsilon_3$ & $\varepsilon_4$ & $\varepsilon_5$& $\varepsilon_6$  & 
$U_{3},\, U_{4}$ & $U_{5},\, U_{6}$ & $K_{3},\, K_{4}$ & $K_{5},\, K_{6}$ & $R_{\alpha 3},\, R_{\alpha 4}$ & $R_{\alpha 5},\, R_{\alpha 6}$ & 
$U'_{34}$ & $U'_{56}$ & $U'_{35,\, 36,\,45,\,46}$ \\ \hhline{|=|=|=|=|=|=|=|=|=|=|=|=|=|=|}
Albiez \textit{et al.}\cite{Albiez2005} & $131.0$ & 133.0 & --  & -- & 0.075 & -- & 0.055  & -- & $\pm 0.063$ & -- & 0.075  & --  &  --  \\ \hline
LeBlanc \textit{et al.}\cite{LeBlanc2011} & $189.0$ & 191.0 & 381.0 & 383.0  & 0.56        & 0.48        & 0.33        & 0.25        & $\pm 0.43$         & $\pm 0.19$         & 0.56      & 0.48      & 0.29     \\               
\hline
\end{tabular}
\caption{Model parameters involving at least one excited trap mode, $n\geq 3 $. 
Note that $\varepsilon_{4,\,6} = \varepsilon_{3,\,5} + 2|J|$.}
\label{tab:multi-mode_params}
\vspace*{0.3cm}
\end{table}

\twocolumngrid

\onecolumngrid

\begin{figure}[t]
\centering
\vspace{-2.2cm}
\subfigure{
	\label{fig:imb_200}
    \begin{minipage}[c][12cm][t]{.48\textwidth}
  \vspace*{\fill}
  \centering
  \includegraphics[width=\textwidth]{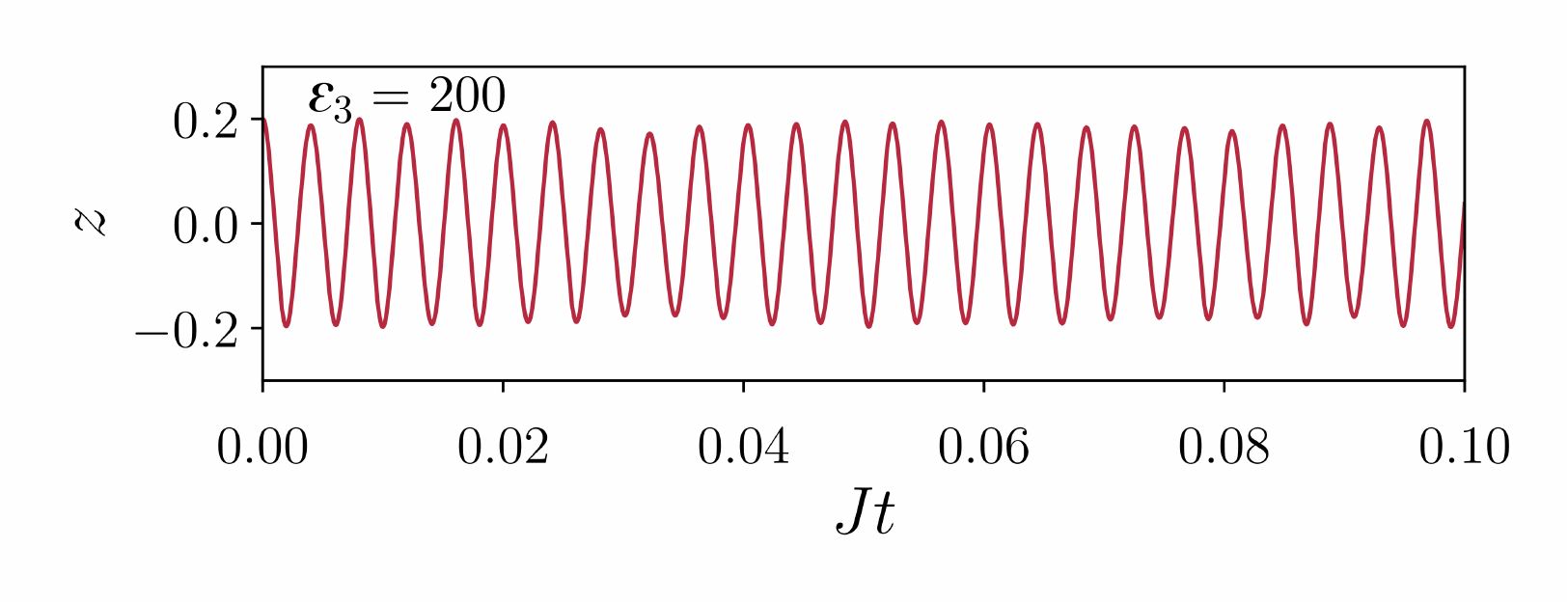}
  \label{fig:BEC_imbalance1}
  \vspace{-1.5cm}
  \includegraphics[width=\textwidth]{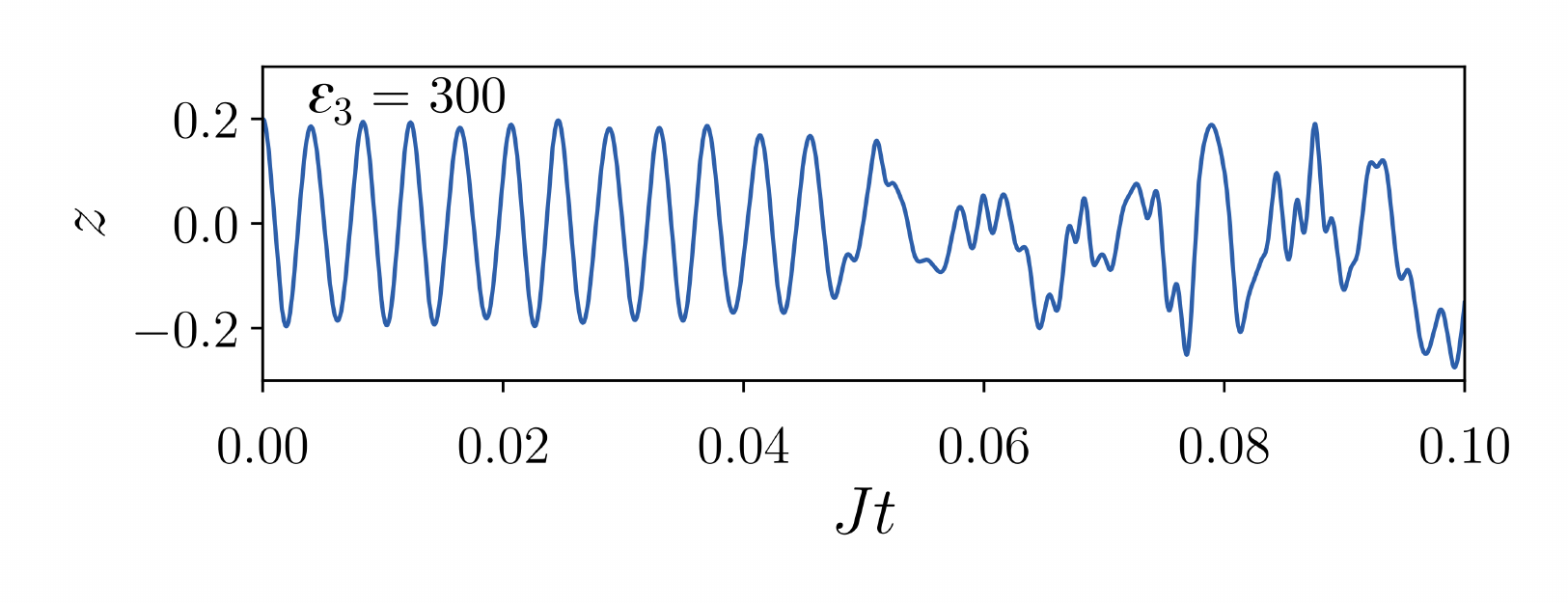}
  \label{fig:BEC_imbalance2}
    \vspace{-1.5cm}
  \includegraphics[width=\textwidth]{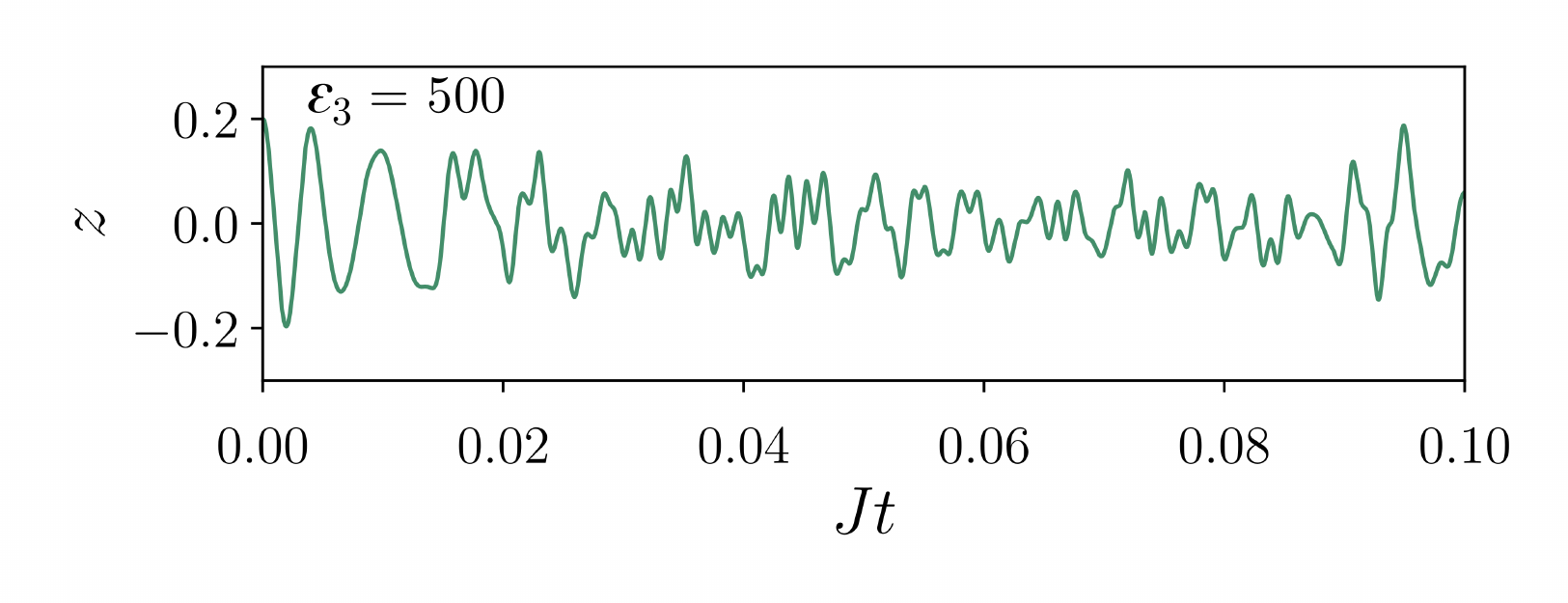}
  \label{fig:BEC_imbalance3}
    \vspace{-1.5cm}
  \includegraphics[width=\textwidth]{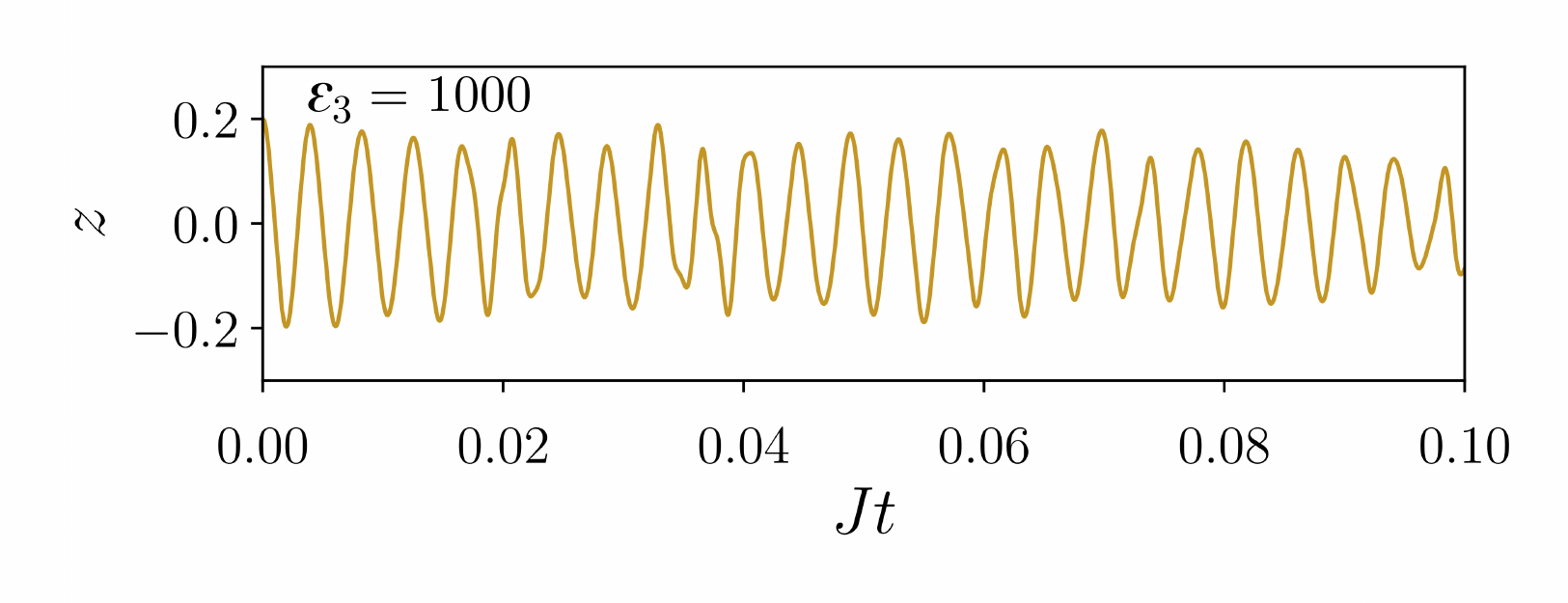}
  \label{fig:BEC_imbalance4}
	\end{minipage}
}
\subfigure{
	\label{fig:flucsToy}
    \begin{minipage}[c][12cm][t]{.48\textwidth}
  \vspace*{\fill}
  \centering
  \includegraphics[width=\textwidth]{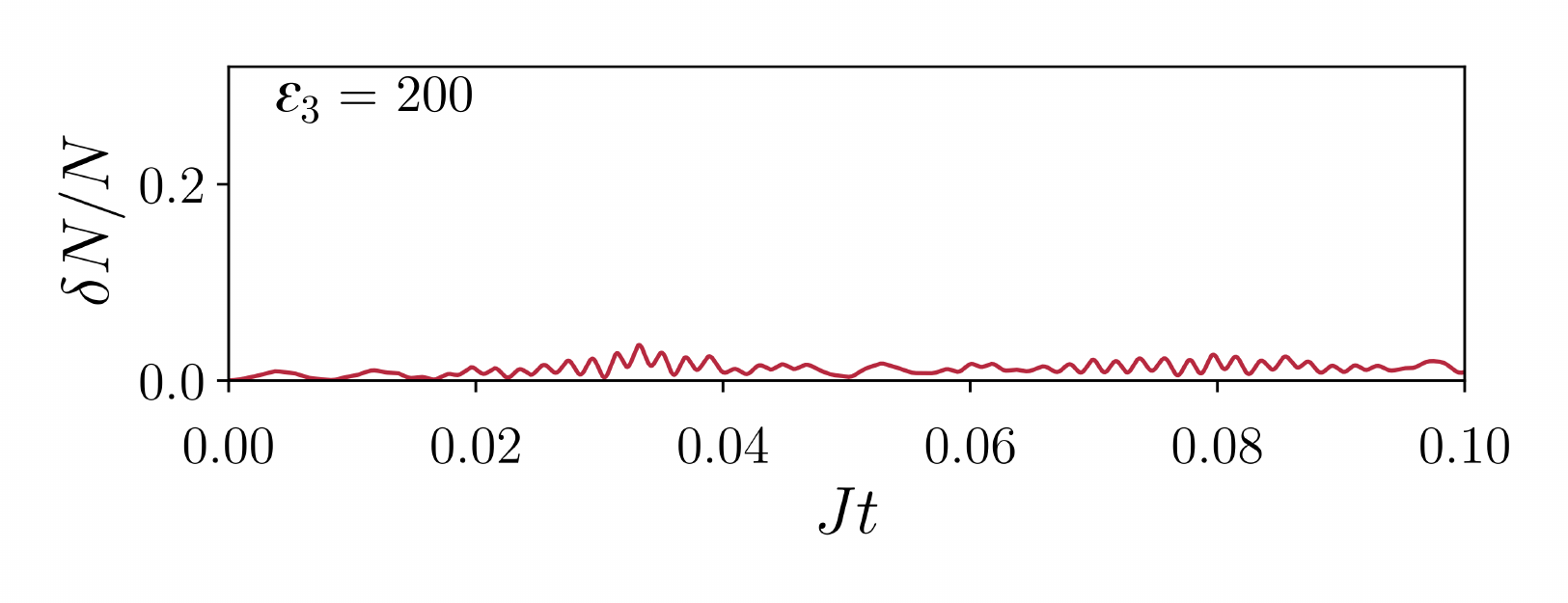}
  \label{fig:fluct1}
  \vspace{-1.5cm}
  \includegraphics[width=\textwidth]{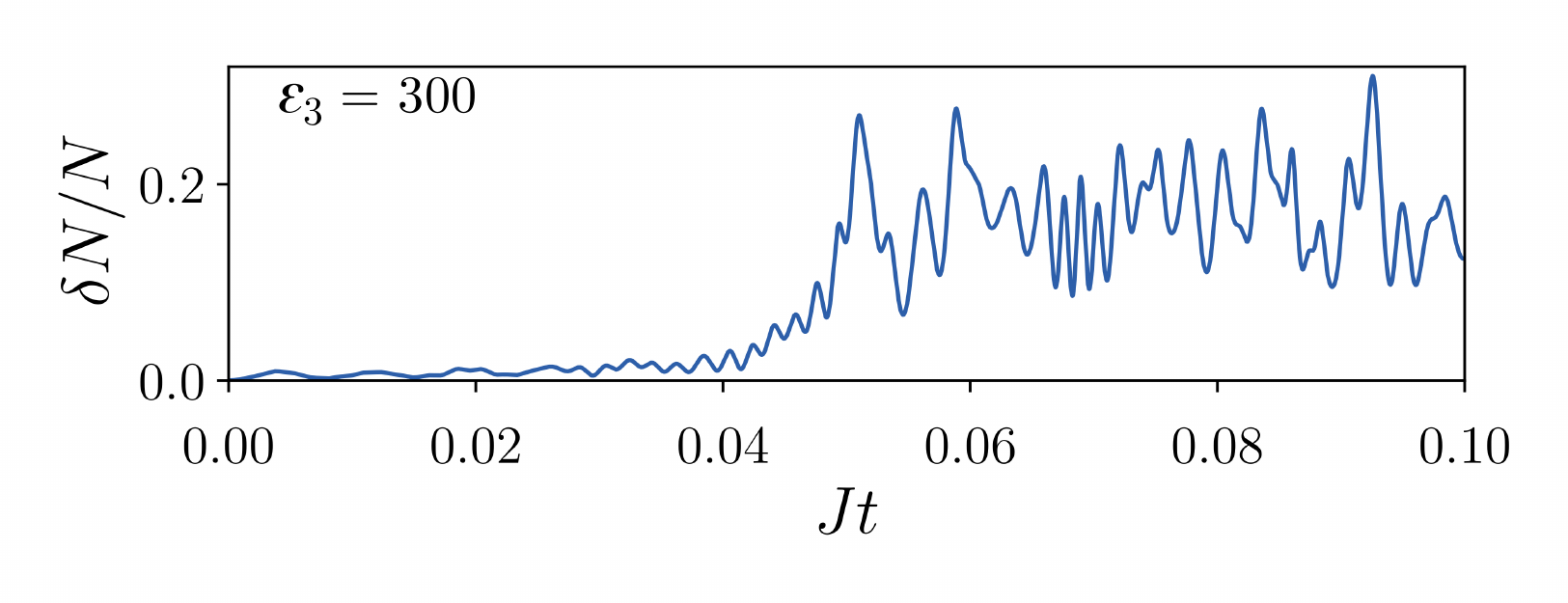}
  \label{fig:fluct2}
    \vspace{-1.5cm}
  \includegraphics[width=\textwidth]{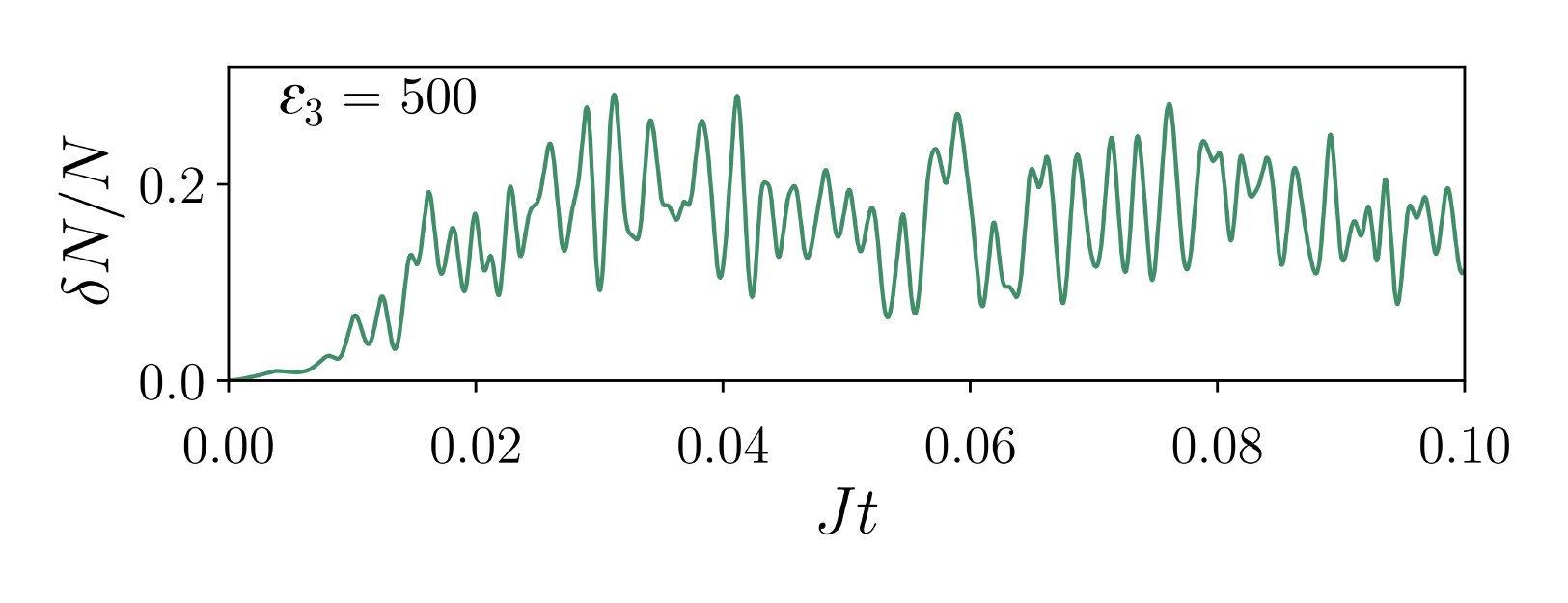}
  \label{fig:fluct3}
    \vspace{-1.5cm}
  \includegraphics[width=\textwidth]{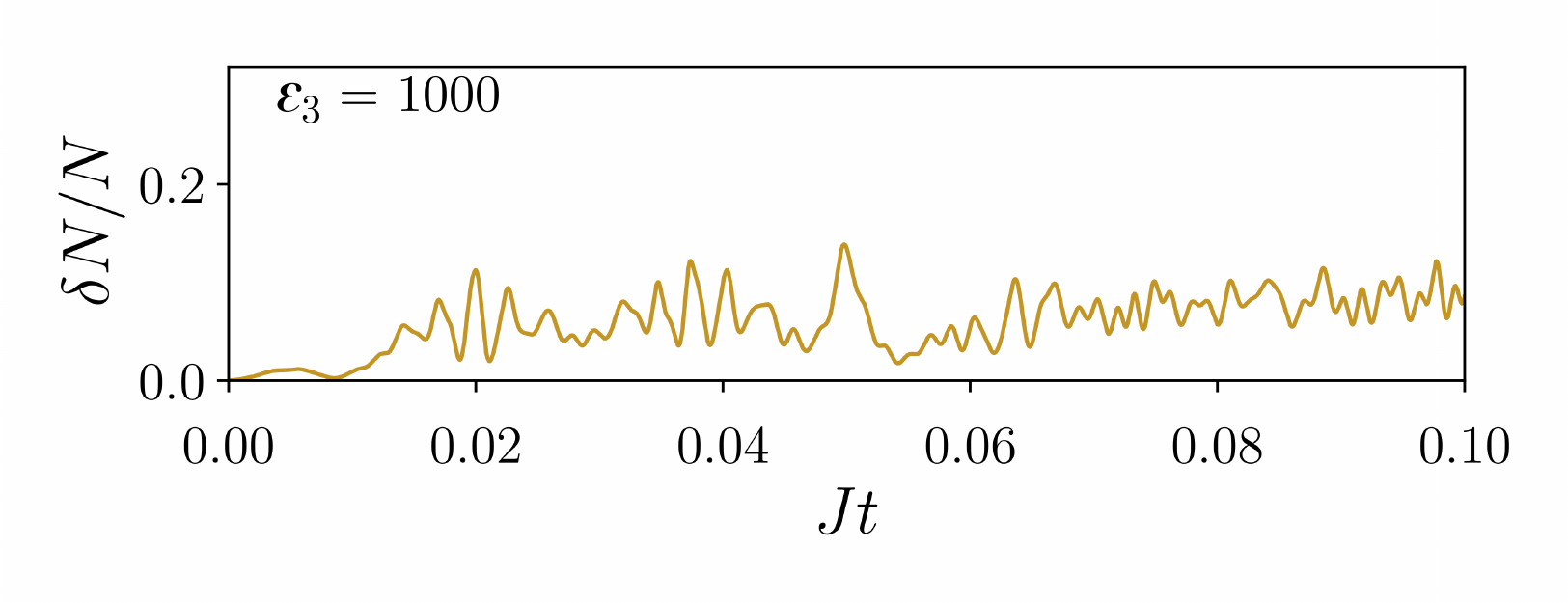}
  \label{fig:fluct4}
	\end{minipage}
} 
\vspace{-0.7cm}
\caption{Population imbalance $z(t)$ (left panels) and total fraction of fluctuations $\delta N(t)/N$ (right panels) for $M = 3$ modes with $N = 5000$ particles, $z(0)=0.2$, initial BEC phase difference $\Delta\theta(0)=0$ and $N_{3}(0) = 0$. In all plots, the parameters are $U = 1.0$, $U'=0$, $J'=-0.05$, $U_3 = 0.5$ $K_3 = 0.1$,  $R_{\alpha 3} = 0.01$, in units of $|J|$. $\varepsilon_3$ is varied and has values $\varepsilon_3 = 200, 300, 500, 1000$ (top to bottom), as indicated.
}
\label{fig:time_traces}
\end{figure}

\twocolumngrid

The trap potentials of the two experiments (A) and (B) have different shapes, 
$V_A({\bf r})$ and $V_B({\bf r})$, respectively, as given in Appendix B.
In order to develop a quantitative description of the dynamics, we  
solve the (noninteracting) Schr\"odinger equation with the 
potentials $V_A({\bf r})$ and $V_B({\bf r})$ for the first ten single-particle 
trap wave functions $\varphi_i({\bf r})$ and compute the matrix elements 
of the full Hamiltonian in the trap eigenbasis according to 
Eqs.~\eqref{eq:eps}-\eqref{eq:U}. 
The general interaction matrix elements 
$U_{ijkl}$ can be classified into intra-level interactions 
($U_{\alpha\alpha\alpha\alpha} \equiv U$, $U_{nnnn} \equiv U_n$),
density-density interactions between different levels 
($U_{\alpha\beta\beta\alpha}\equiv U'$, $U_{n m m n}  \equiv U_{nm}'$, 
$U_{\alpha m m \alpha}\equiv K_{m}$) 
and interaction-induced transitions between different levels
($U_{\alpha\alpha\alpha\beta}\equiv J'$, 
$U_{\alpha\alpha\alpha m}\equiv R_{\alpha m}$). Here $\alpha,\beta =1,\,2$, 
$\alpha \neq \beta$, denote the 
ground modes $\varphi_{1, 2}({\bf r})$ localized in the left or right potential well and $n,\,m=3,\,4,\,5,\,\dots$ 
the higher trap levels. 
See Appendix B for details of the definitions and calculations.
The parameter values computed for a $^{87}$Rb gas (scattering 
length $a_s\approx 98\, a_0$)\cite{Vogels1997} in the experimental setups
(A) and (B) are listed in Tabs.~\ref{tab:two-mode_params} 
and \ref{tab:multi-mode_params}. 
The {\it bare} Josephson coupling $J$ turns out to be approximately equal 
for both experiments, (A) and (B),  
$J \approx -2\pi \times 0.16 \;\mathrm{Hz}$ (see Eq. \eqref{eq:J} and 
Appendix B). All energies in this paper are given in units of $|J|$.

\vspace*{0.1cm}

\subsection{Resonant single-particle excitations}
\label{subsec:resonant}

\begin{figure}[b]
  \includegraphics[width=0.5\textwidth]{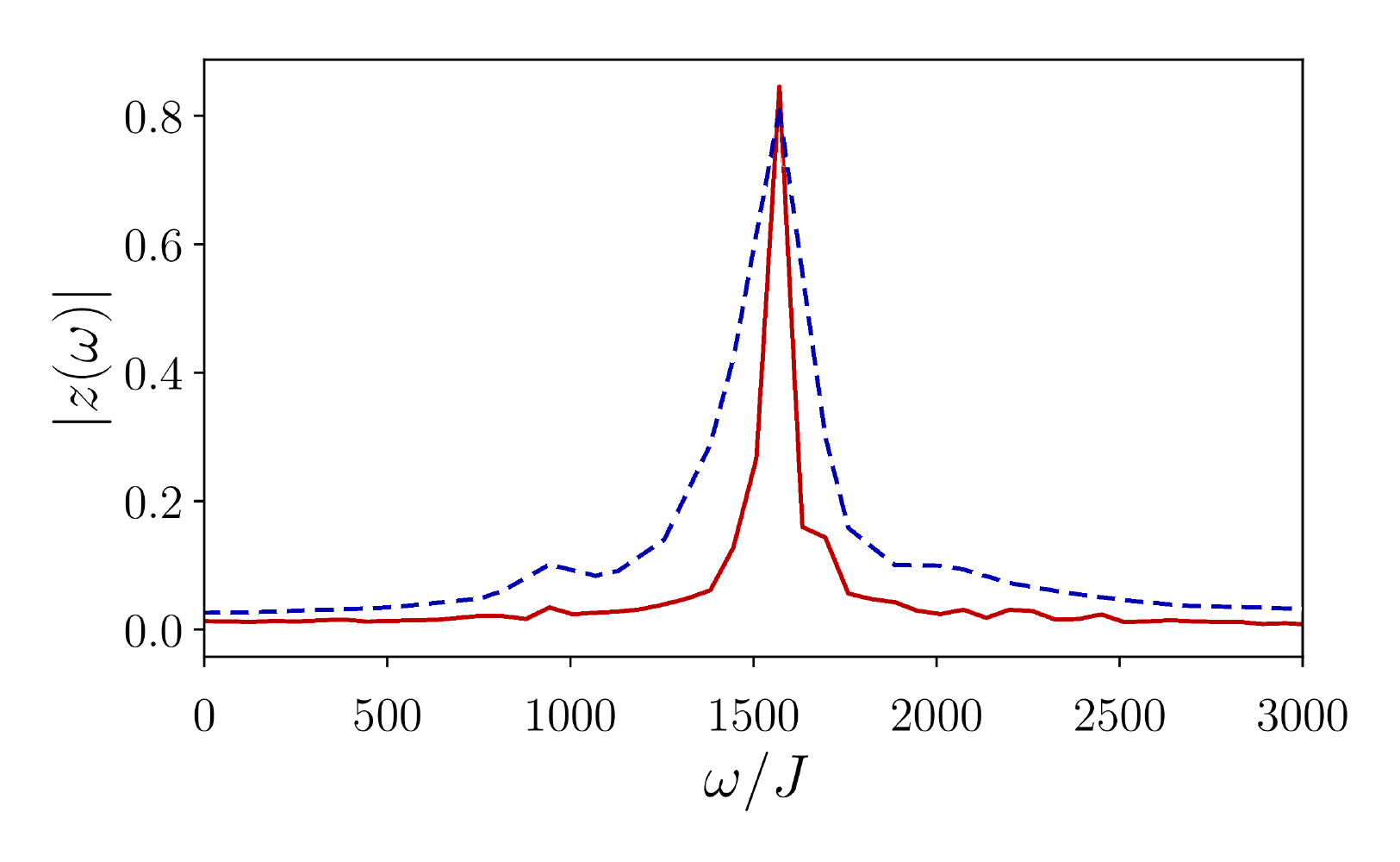}
  \vspace{-1cm}
  \caption{Magnitude spectrum (absolute value of the Fourier transform) of the population imbalance $z(t)$ for the parameters given in Fig. \ref{fig:scan} for $\varepsilon_3 = 200$ (solid line) and $\varepsilon_3 = 300$ (dashed line), $N=5000$. 
The time interval of the FFT was truncated at the onset of the fluctuation regime: for $\varepsilon_3 = 300$, the time trace was cut at $Jt = 0.04$, whereas for $\varepsilon_3 = 200$, the entire displayed interval was used.}\label{fig:fft_200_300}
\end{figure}

In this subsection, we establish that incoherent excitations 
(fluctuations) out of the condensate are efficiently created, and therefore that
damping occurs, if the frequency of the Josephson oscillations is 
in resonance with one of the renormalized single-particle levels. 
The interactions not only renormalize the single-particle levels, but 
also the Josephson frequency, $\omega_J\to\tilde\omega_J$. 
Within the two-mode model in the linear regime of 
Josephson oscillations, it is given by\cite{Smerzi1997} 
\begin{equation}
\tilde{\omega}_J = 2J\sqrt{1+NU/2J}\ .
\label{eq:omega_J}
\end{equation} 
In the general case of multiple modes and inter-mode interactions it is,
however, not possible to give an analytical expression. Therefore, we
numerically evolve the interacting system in time for a large number 
of oscillations, using the Keldysh equation-of-motion method presented in section \ref{sec:formalism}, and extract the 
renormalized Josephson frequency $\tilde\omega_J$ 
from the Fourier spectrum of the time-dependent BEC population 
imbalance $z(t)$. 
To establish the resonance condition for realistic experimental setups, 
we consider an exemplary  
system of three modes with typical parameter values for the 
experiments (A), (B), given in the caption of Fig.~\ref{fig:time_traces}, 
and vary the bare energy $\varepsilon_3$ of the third mode above 
the two lowest modes, whose bare energy we set to
$\varepsilon_1=\varepsilon_2 = 0$. 
The corresponding time traces of the BEC population 
imbalance $z(t)=[N_1(t)-N_2(t)]/N$ and of the total fraction 
of noncondensed particles $\delta N(t)/N$ (fluctuations) are shown in 
Fig.~\ref{fig:time_traces}. For small and for large 
level spacings, $\varepsilon_3=200,\ 1000$, essentially no 
fluctuations are generated (right panels), 
and the Josephson oscillations remain undamped (left panels). 
However, for intermediate level spacings, 
$\varepsilon_3=300$, and more so for $\varepsilon_3=500$, we observe
efficient excitation of fluctuations at a characteristic time 
$\tau_c$, and at the same time scale the oscillations become 
depleted and irregular, but remain reproducible.\cite{Trujillo-Martinez2009}
Inelastic interactions between these incoherent excitations 
(not taken into account at the BHF level of approximation) will lead to 
rapid damping and eventual thermalization of the Josephson oscillations, 
as shown in Ref.~[\onlinecite{Posazhennikova2016}]. 
To determine the Josephson frequency of the interacting system, $\tilde\omega$,
we compute the magnitude spectrum of $z(t)$ by fast Fourier transform (FFT) 
of the time traces up to the time $\tau_c$, i.e., before fluctuations are  
efficiently generated, as shown in Fig.~\ref{fig:fft_200_300}. 
$\tilde\omega_J$ is given by the position of the pronounced peak in these 
spectra.

\begin{figure}[t]
  \includegraphics[width=0.5\textwidth]{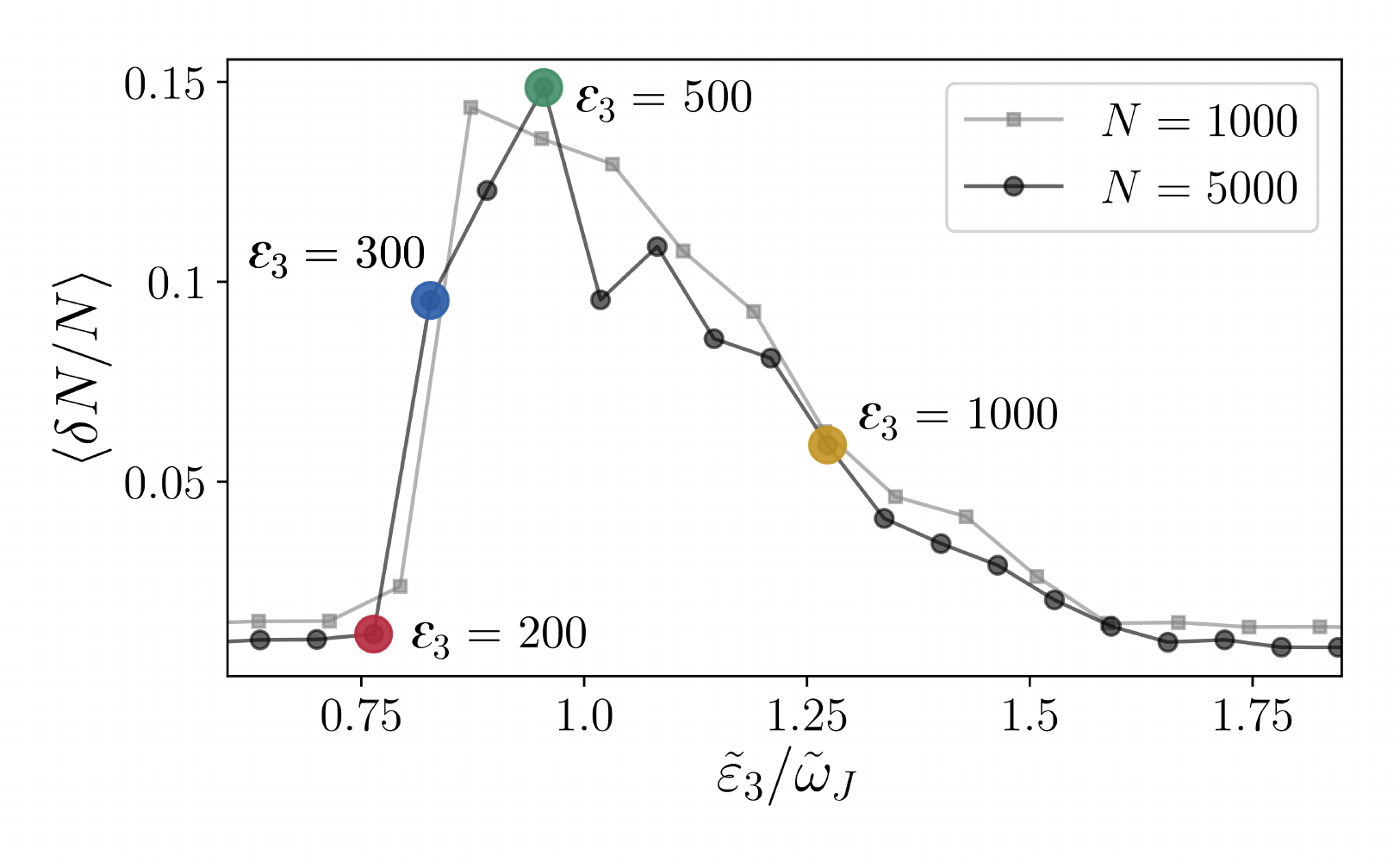}
  \vspace{-1cm}
  \caption{Time-averaged fraction of fluctuations as a function of the ratio of the effective single-particle energy $\tilde{\varepsilon}_3$ and the effective Josephson frequency $\tilde{\omega}_J$ for two different particle numbers (the small, gray squares are for $N=1000$, the small, black circles are for $N=5000$, corresponding to experiments (A) and (B), respectively). The interaction parameters are $U = 1.0$, $U'=0$, $J'=-0.05$, $U_3 = 0.5$, $K_3 = 0.1$,  $R_{\alpha 3} = 0.01$,
and the initial conditions: initial population imbalance $z(0)=0.2$, initial phase difference $\Delta \theta(0)=0$, $M = 3$, and $N_{3}(0) = 0$. 
The renormalized Josephson frequency is extracted from the Fourier transforms of $z(t)$: $\tilde{\omega}_J \approx 1571$ for $N=5000$, and  $\tilde{\omega}_J \approx 315$ for $N = 1000$. The time average of the fluctuations was taken over the displayed time interval for $N = 5000$, whereas for $N = 1000 $ an interval of 5 times that size was used. The thick dots represent the results 
for the time traces of Fig.~\ref{fig:time_traces} for the corresponding 
values of $\varepsilon_3$, as indicated in the figures.}\label{fig:scan}
\end{figure}

To analyze now the fluctuation excitation mechanism quantitatively, 
the renormalized level $\tilde\varepsilon_3$ as well as the time traces $z(t)$,
$\delta N(t)/N$ are computed for a large number of bare $\varepsilon_3$ 
values, and the time-averaged fraction of fluctuations, $\langle \delta N/N\rangle$, is plotted as a function of the 
ratio $\tilde\varepsilon_3/\tilde\omega_J$, see Fig.~\ref{fig:scan}. 
The figure clearly exhibits resonant behavior: 
The fluctuation fraction reaches a broad but pronounced maximum when 
the renormalized level $\tilde\varepsilon_3$ and the renormalized Josephson
frequency $\tilde\omega_J$ coincide.  

We note that this resonant fluctuation-creation mechanism is closely related to, but more general than the dynamical mean-field instabilities reported in Refs.~[\onlinecite{Castin1997, Vardi2001,Gerving2012}]. 
It leads to the highly nonlinear, abrupt creation creation of
fluctuations\cite{Trujillo-Martinez2009} 
at the characteristic time $\tau_c$ seen, e.g., 
in Fig.~\ref{fig:time_traces}, panels of the second row ($\varepsilon_3 =300$). 
The frequency $\tilde {\omega}_J$ acts like the frequency of an external driving field for the subsystem of non-condensate excitations (fluctuations). However, in the present Josephson system the driving is an intrinsic effect, not an external one as in Ref.~[\onlinecite{Castin1997}]. 
Also, our approach is not restricted
to the two-mode scenario,\cite{Vardi2001} but can be extended to any  
number of modes involved. We have tested this for various sets of 
parameter values and system sizes.

Incoherent excitations will lead to rapid damping and eventual thermalization  of the system.\cite{Posazhennikova2016} In order to avoid damping and to 
stabilize coherent motion, one needs to tune the away from the resonance. 
One way of achieving this is to change the particle number $N$: 
While the excitation energies of the {\it not macroscopically occupied} levels,
$\tilde\varepsilon_{n}$, $n\geq 3$, are not strongly affected by $N$,
$\tilde \omega_J$ depends sensitively on $N$ [c.f.~Eq.\eqref{eq:omega_J}],
so that the resonance condition $\tilde\omega_J\approx\tilde\varepsilon_{n}$
(c.f. Fig.~\ref{fig:scan}) may easily be avoided.

\subsection{Comparison with experiments}
\label{subsec:comparison}

We now examine how the experiments (A) and (B) fit into the resonant-fluctuation-creation scenario described above. 

\begin{table}[b]
\centering
\begin{tabular}{|c|c|c|c|c|c|c|c|c|}
\hline 
   & $N$  & $z(0)$   & $N_1$ & $N_2$ & $N_3$ & $N_4$ & $N_5$ & $N_6$ \\ 
\hhline{|=|=|=|=|=|=|=|=|=|}
Albiez \textit{et al.}\cite{Albiez2005} & 
     1150 & 0.290 & 742   & 408   & 0     & 0     &  --     &  --   \\
\hline
LeBlanc \textit{et al.}\cite{LeBlanc2011} & 4500 & 0.116 & 2436  & 1914  & 75    & 75    & 0                     & 0\\                   
\hline
\end{tabular}
\caption{Occupation numbers at time $t=0$ used for the 
numerical calculations.}
\label{tab:occup_params}
\end{table}

\begin{figure}[t]
\includegraphics[width=\linewidth]{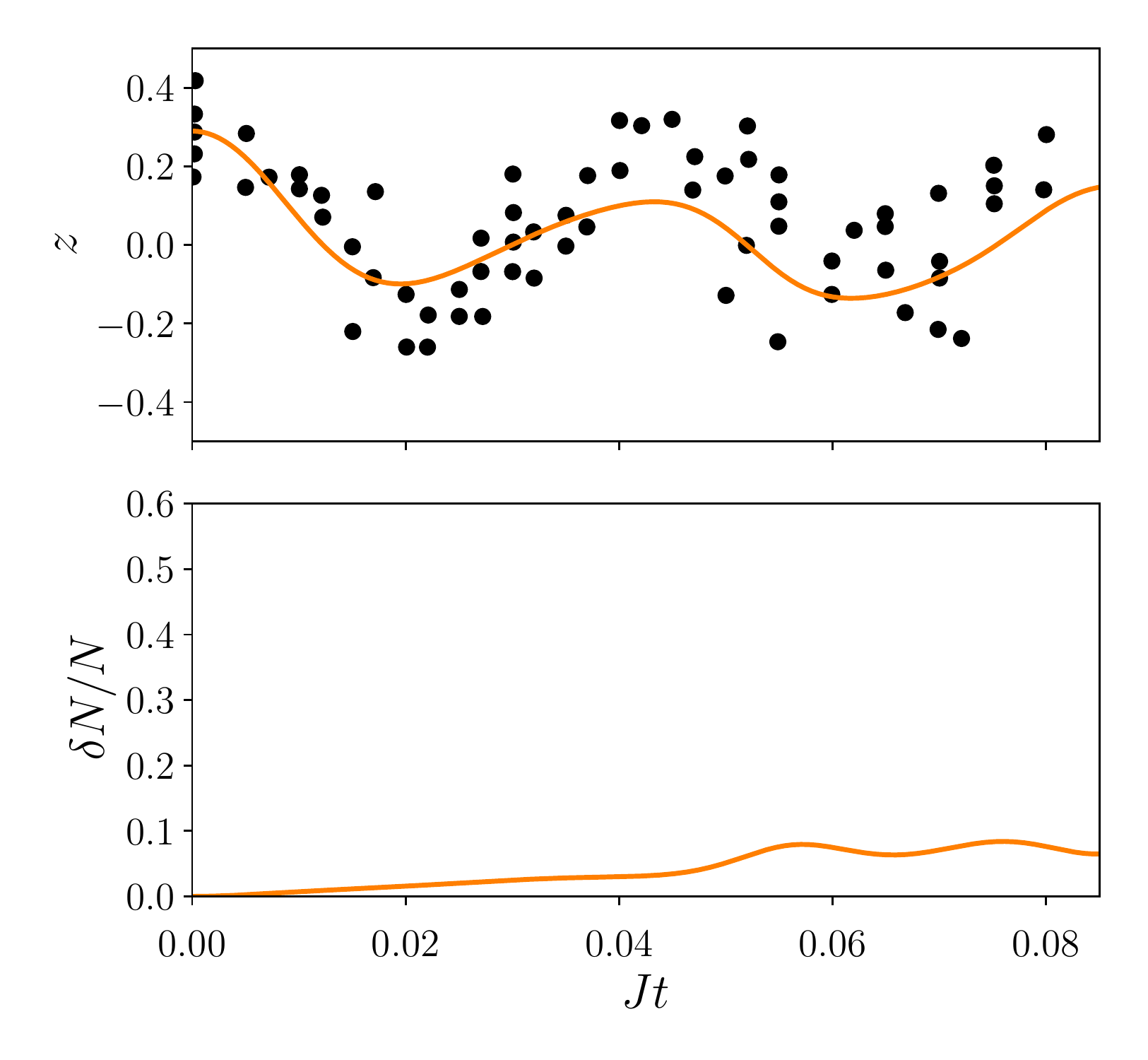}
\vspace{-1.0cm}
\caption{Population imbalance $z(t)$ and relative fraction of fluctuations 
$\delta N(t)/N$ for the experiments (A). \cite{Albiez2005} 
The experimental data points (black dots) are taken from the reference. 
The parameters and initial conditions for the calculations 
are listed in Tabs.~\ref{tab:two-mode_params} -- \ref{tab:occup_params}.}\label{fig:obiImb}
\end{figure}

In Fig.~\ref{fig:obiImb} we give the results of our calculations for the experimental setup (A) of Albiez {\it et al.} \cite{Albiez2005} with the four relevant modes shown in Fig.~\ref{fig:levels} in direct comparison with the experimental data points. Note that there is no fitting of parameters involved. We see that the agreement with the experiment is very good regarding both the frequency and the amplitude of the Josephson oscillations. 
In particular, no damping is observed in the experiment as well as in the 
calculation. The fraction of fluctuations remains below 10 $\%$, indicating 
that this experimental setup is away from the resonance discussed 
in Fig. \ref{fig:scan}. 

\begin{figure}[t]
\includegraphics[width=0.482\textwidth]{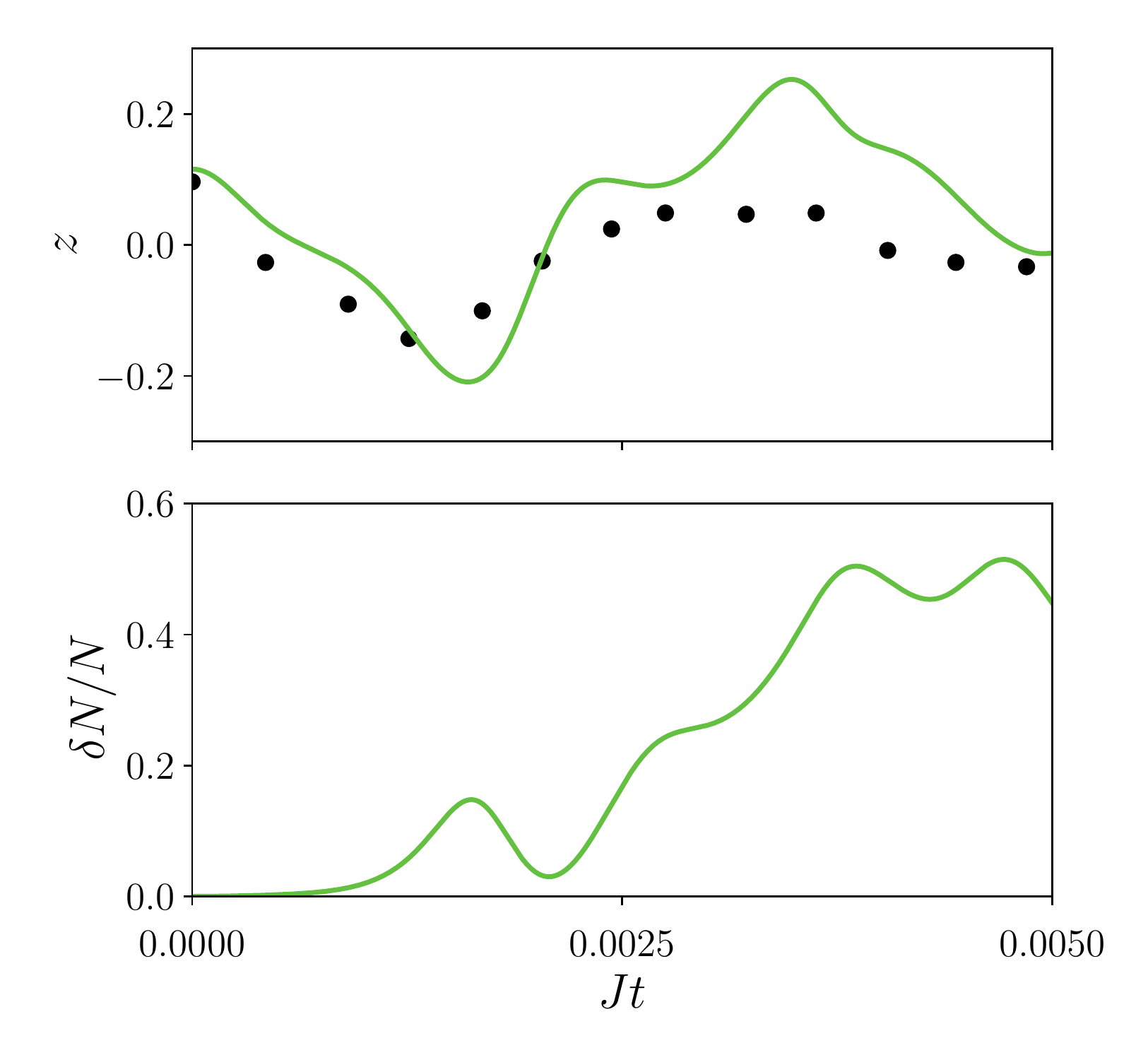}
\vspace{-1.0cm}
\caption{Population imbalance $z(t)$ and relative fraction of fluctuations 
$\delta N(t)/N$ for the experiments (B). \cite{LeBlanc2011} 
The experimental data points (black dots) are taken from the reference. 
The parameters and initial conditions for the calculations 
are listed in Tabs.~\ref{tab:two-mode_params} -- \ref{tab:occup_params}.}
\label{fig:thyImb}
\end{figure}

In Fig. \ref{fig:thyImb} we display the corresponding calculations for the 
experiment (B).\cite{LeBlanc2011} We took six relevant modes into account in our calculations, as explained in the discussion 
of Fig. \ref{fig:levels}. For this experiment we assume a small initial 
condensate occupation of the modes $m=3,\,4$, as listed in 
Tab.~\ref{tab:occup_params}, because of the small excitation 
energy of these modes (see Fig.~\ref{fig:levels}) with regard to the larger 
interaction parameters of experiment (B). 
Here the agreement with experiment is quantitatively not as good as for 
the experiment (A).\cite{Albiez2005} However, the theoretical calculation 
reproduces the strong amplitude reduction of $z(t)$ after a short time of 
only $t\approx 0.004~J$ in agreement with experiment. At the same time,
the calculation shows a fast and efficient excitation of fluctuations,
which set in at a characteristic time scale\cite{Posazhennikova2016} 
of $\tau_c\approx 0.0013~J$ and reach a maximum amplitude of about
$\delta N_{max} /N \approx 0.5$ near the time $t\approx 0.0035~J$. This
indicates that this experimental setup is in the resonant regime. 
Importantly, we find that the efficient creation of fluctuations
for the parameters of experiment (B) is robust, independent of the small 
condensate occupation of the modes with $m=3,\,4$ as well as the precise 
value of $N$. 

The reason for the reduced quantitative agreement with experiment can be 
understood from the behavior of the fluctuation fraction.  
As seen in Fig. \ref{fig:thyImb}, lower panel, the departure of the theoretical 
results from the experimental data points is significant for those times when 
the non-condensate fraction $\delta N(t)/N$ is large. 
A large fraction of fluctuations means that the BHF approximation employed 
in the present work is not sufficient, and higher-order corrections should be
taken into account. They account for inelastic collisions of excitations and 
will, therefore, lead to rapid damping,\cite{Posazhennikova2016} as 
observed in experiment (B).\cite{LeBlanc2011}

\section{Discussion and conclusion}\label{sec:discussion}

We have considered Josephson oscillations of isolated, atomic BECs trapped in double-well potentials and analyzed the impact of fluctuations, i.e. out-of-condensate particle excitations, on the dynamics of the oscillations for the two specific experiments  of Albiez {\it et al. } (A),\cite{Albiez2005} and  of LeBlanc {\it et al.} (B).\cite{LeBlanc2011} While the first experiment is
well described by Gross-Pitaevskii dynamics,\cite{Smerzi1997}
suggesting a negligible role played by the fluctuations, the latter experiment 
exhibits fast relaxation of the oscillations, which is not contained in the 
semiclassical Gross-Pitaevskii description, even if multiple trap modes 
are considered. One therefore expects a sizable number of non-condensate 
excitations created in this experiment.

We identified a scenario for the resonant excitation of fluctuations. It indicates that, whenever any of the renormalized trap levels is close to the effective Josephson frequency, this leads to resonant creation of fluctuations 
and a departure from the Gross-Pitaevskii dynamics. The interaction-induced 
renormalization of both the trap levels as well as the Josephson frequency
is important for this resonant effect to occur. 
By numerical calculations for the realistic model parameters, we showed that 
indeed experiment (A) is off resonance with only a small amount of 
fluctuations created, while experiment (B) 
is operated in the resonant regime and dominated by fluctuations. 
This reconciles the qualitatively different behavior of the two experiments.  
In another, more recent experiment\cite{Spagnolli2017} the bare Josephson 
frequency $\omega_J$ was chosen smaller than the trap level spacings 
(see Supplemental Information to Ref.~[\onlinecite{Spagnolli2017}]), 
and the BJJ oscillation frequency was further reduced by tuning the 
interaction $U$ to become attractive. Thus, this experiment is in the 
off-resonant regime. Indeed, it shows extended undamped oscillations. 
It is well described by GPE dynamics alone\cite{Spagnolli2017}, as expected.

As a more general conclusion, for the design of long-lived, coherent 
Josephson junctions it is essential to ensure that none of the renormalized
and possibly interaction-broadened trap levels is on resonance with the 
effective Josephson frequency. This can be achieved by either tuning the 
parameters of the trap or by adjusting the total number of particles. 
In this way, Bose-Josephson junctions may serve as a device for studying 
the departure from classicality due to quantum fluctuations in a controlled way.

\section*{ACKNOWLEDGMENTS} We would like to thank A. Nejati, B. Havers and 
M. Lenk for useful discussions and especially  J. H. Thywissen and
L. J. LeBlanc for providing us with the details of their trapping potential.
This work was supported by the Deutsche Forschungsgemeinschaft (DFG) 
through SFB/TR 185.\\

\bibliographystyle{apsrev4-1}
\bibliography{refs}

\newpage

\section*{APPENDIX A: TWO-MODE APPROXIMATION}

To illustrate the details of the formalism, we present here the derivation of
the equations of motion for a two-mode system where, however, the 
out-of-condensate fluctuations are taken into account in each mode. In this 
respect, the calculation goes beyond the two-mode model studied at 
the semiclassical (Gross-Pitaevskii) level of approximation in 
Refs.~[\onlinecite{Smerzi1997, Ananikian2006}]. 
For clarity of presentation, we here discard the 
nonlocal (inter-mode) interaction parameters. 
The important steps to be demonstrated in this appendix carry over to the general case used to describe the experiments (multi-mode, nonlocal interactions) in a straightforward manner. For the scope of this appendix, the action hence reads,
\[S=S_{0}+\sum_{\alpha=1}^{2}S_{\text{int}}[\phi_{\alpha}^{*},\phi_{\alpha}^{}],\]
where
\begin{equation}
S_{0}=\int \text{d}t\left[\sum_{\alpha=1}^{2}\left(\phi_{\alpha}^{*}G_{0}^{-1}\phi_{\alpha}^{}\right) - J\left( \phi_{1}^{*}\phi_{2}^{} + \phi_{2}^{*}\phi_{1}^{} \right)\right],
\end{equation}
and
\begin{equation}
S_{\text{int}}[\phi^{*},\phi]=-\tfrac{U}{2}\int \text{d}t\,\left|\phi\right|^4.
\end{equation}
Writing the corresponding Keldysh action explicitly, one finds
{\small
\begin{align}
  \begin{split}
    & S_{K}[\Phi^{}_{c},\Phi^{}_{q}]=\int \text{d}t\bigg\{\sum_{\alpha=1}^{2}\left[\right.\phi^{*}_{\alpha\,q} \left( \ii\,\partial_t - \varepsilon \right) \phi_{\alpha\,c}^{} \\
    & +\phi^{*}_{\alpha\,c} \left( \ii\,\partial_t - \varepsilon \right)  \phi_{\alpha\,q}^{}\left.\right]  - J\left[ \phi_{1\,q}^{*}\phi_{2\,c}^{} + \phi_{1\,c}^{*}\phi_{2\,q}^{}  + \text{c.c.}\right] \\
    & -\tfrac{U}{2} \sum_{\alpha=1}^{2}\left[\phi_{\alpha\,c}^{*}\phi_{\alpha\,c}^{*}\phi^{}_{\alpha\,c}\phi^{}_{\alpha\,q}+
       \phi^{*}_{\alpha\,q} \phi^{*}_{\alpha\,q}\phi^{}_{\alpha\,q}\phi^{}_{\alpha\,c} + \text{c.c.}\right]\bigg\}.
  \end{split}
\end{align}
}%
Performing the variation according to Eq. \eqref{eq:SPAc} yields the modified Gross-Pitaevskii equation (GPE) as the saddle-point equation of our action:
\begin{align}
\begin{split}\label{eq:SPAc1}
\ii\partial_t \Phi^{}_{1\,c}&=\varepsilon \Phi^{}_{1\,c}+J\Phi^{}_{2\,c}+\tfrac{U}{2}\Phi^{*}_{1\,c}\Phi^{}_{1\,c}\Phi^{}_{1\,c}\\
&+\tfrac{U}{2}(\Phi^{}_{1\,q}\Phi^{}_{1\,q}\Phi^{*}_{1\,c}+2\Phi^{*}_{1\,q}\Phi^{}_{1\,q}\Phi^{}_{1\,c}\\
&+2\langle \delta\phi^{}_{1\,c}\delta\phi^{*}_{1\,c} \rangle\Phi^{}_{1\,c}+\langle\delta\phi^{}_{1\,c}\delta\phi^{}_{1\,c}\rangle\Phi^{*}_{1\,c}\\
&+2\langle\delta\phi^{}_{1\,c}\delta\phi^{}_{1\,q}\rangle\Phi^{*}_{1\,q}+2(\langle\delta\phi^{}_{1\,c}\delta\phi^{*}_{1\,q}\rangle+\text{c.c.})\Phi^{}_{1\,q}\\
&+2\langle\delta\phi^{}_{1\,q}\delta\phi^{*}_{1\,q}\rangle\Phi^{}_{1\,c}+\langle\delta\phi^{}_{1\,q}\delta\phi^{}_{1\,q}\rangle\Phi^{*}_{1\,c}).
\end{split}
\end{align}
Taking into account that $\Phi_{1\,q}=\Phi^{*}_{1\,q}=0$, as well as the fact that all Green functions of two quantum fields vanish because of the relation between (anti-) time-ordered, greater and lesser Green functions, by letting $\Phi_{\alpha\,c}=\Phi^{}_{\alpha}$ we obtain the final form of our modified GPE as
\begin{align}
\ii\partial_t \Phi^{}_{1}&=\varepsilon \Phi^{}_{1}+J\Phi^{}_{2}+\tfrac{U}{2}\Phi^{*}_{1}\Phi^{}_{1}\Phi^{}_{1},\nonumber\\
&+\tfrac{U}{2}(\langle\delta\phi^{}_{1\,c}\delta\phi^{}_{1\,c}\rangle\Phi^{*}_{1}+2\langle \delta\phi^{}_{1\,c}\delta\phi^{*}_{1\,c}\rangle\Phi^{}_{1}),
\end{align}
which upon introduction of the fluctuation Green functions reads
\begin{align}\label{eq:GPE}
\ii\partial_{t}\Phi^{}_{1}=&(\varepsilon+\tfrac{U}{2}\Phi^{*}_{1}\Phi^{}_{1})\Phi^{}_{1}+J\Phi^{}_{2}+ \ii U(\,G_{11}\,\Phi^{}_{1}+\tfrac{1}{2}\,g_{11}\,\Phi^{*}_{1}).
\end{align}
The equation for the second field can be obtained by substituting $2\leftarrow 1$ and vice versa. Next we calculate the second derivatives of the effective action and find
\begin{align}
\frac{\delta^{2}\Gamma}{\delta\Phi_{\alpha\,q}^{*}(t)\delta\Phi_{\alpha\,c}^{}(t)}&=\ii\partial_t -\varepsilon - U\left(\Phi^{*}_{\alpha}\Phi^{}_{\alpha} + \ii G_{\alpha\alpha}\right),\\
\frac{\delta^{2}\Gamma}{\delta\Phi_{\alpha\,q}^{*}(t)\delta\Phi_{\alpha\,c}^{*}(t)}&=-\tfrac{U}{2}\left( \Phi^{\,2}_{\alpha}+ \ii g_{\alpha\alpha} \right),
\end{align}
whereas the off-diagonals in level space are simply
\begin{align}
\frac{\delta^{2}\Gamma}{\delta\Phi_{1\,q}^{*}(t)\delta\Phi_{2\,c}^{}(t)}&=-J,\\
\frac{\delta^{2}\Gamma}{\delta\Phi_{1\,q}^{*}(t)\delta\Phi_{2\,c}^{*}(t)}&=0.
\end{align}

Now make the ansatz
\begin{align}
\Phi^{}_{\alpha}=\sqrt{2N_\alpha}\,e^{\ii\varphi_{\alpha}}
\end{align}
for the condensate fields. Subtracting Eq. \eqref{eq:dyson} and the corresponding advanced equation, and taking the upper left component of the matrices in Bogoliubov space, one finds, after performing the equal-time limit on the Green functions $G_{\alpha\beta}(t,t')$, that
{\small
\begin{align}\label{eq:G_diag}
\begin{split}
 \ii\partial_tG_{11} + N_1 U(e^{-2\ii\varphi_1}{g_{11}}^{} + \text{c.c.}) + J(G_{12} - G_{21})&=0,\\
 \ii\partial_t G_{22} + N_2 U(e^{-2\ii\varphi_2}{g_{22}}^{} + \text{c.c.}) - J(G_{12} - G_{21})&=0,
 \end{split}
\end{align} }%
where $G_{\alpha\beta}=G_{\alpha\beta}(t)=G_{\alpha\beta}(T=t,\tau = 0)$ depends only on the average time $T=(t+t')/2=t$ after taking $t'\to t$. 

The same holds for the anomalous Green functions. Accordingly, adding Eq. \eqref{eq:dyson} and the corresponding advanced equation, and taking the upper right component in Bogoliubov space, one finds for the anomalous Green functions, e.g.
\begin{align}\label{eq:anomG}
 \ii\partial_t g_{\alpha\alpha} &- 2(\varepsilon + 2N_\alpha U + \ii UG_{\alpha\alpha})g_{\alpha\alpha} \nonumber\\
 &- U  \left( 2N_\alpha e^{2\ii\varphi_{\alpha}} + \ii g_{\alpha\alpha} \right) G_{\alpha\alpha} - 2Jg_{12}=0.
\end{align}
The remaining equations are
{\small
\begin{align}\label{eq:G_offDiag}
	\begin{split}
		&\ii\partial_t G_{12} - U\left( 2(N_1-N_2) + \ii G_{11}-\ii G_{22}\right)G_{12} + J(G_{11}-G_{22}) \\
        &+ \tfrac{U}{2}\left[\right. g_{12}^{*}(\Phi_1\Phi_1+\ii g_{11}) + g_{12}(\Phi_2^{*}\Phi_2^{*} -\ii g_{22}^{*}) \left.\right] = 0,
	\end{split}
\end{align}
}%
and
\begin{align}\label{eq:g_offDiag}
	\begin{split}
		&\ii\partial_t g_{12} -  2\varepsilon g_{12} + U\sum_{\alpha}(2N_\alpha+ \ii G_{\alpha\alpha})g_{12}  - J\sum_{\alpha}g_{\alpha\alpha}\\
        &+ \tfrac{U}{2}\left[ G_{12}^{*}(\Phi_1\Phi_1+\ii g_{11}) - G_{12}(\Phi_2^{*}\Phi_2^{*} + \ii g_{22}) \right] = 0,
	\end{split}
\end{align}
together with the identities $G_{21}(t)=-G_{12}^{*}(t)$ and $g_{21}(t)=g_{12}(t)$.

With $G_{\alpha\alpha}=-\ii F_{\alpha}$, where
\begin{align}
F_{\alpha}=2\delta N_\alpha + 1,
\end{align}
one obtains for the total number of fluctuations
\begin{align}\label{eq:flucsNum}
\delta\dot{N}=\delta\dot{N}_1 + \delta\dot{N}_2= -\sum_{\alpha=1}^2\tfrac{N_\alpha U}{2}( e^{-2\ii\varphi_{\alpha}}{g_{\alpha\alpha}}^{} + \text{c.c.}).
\end{align}
Defining the phase difference of the two condensates as $\Delta\varphi=\varphi_2-\varphi_1$, from Eq. \eqref{eq:GPE} one calculates
\begin{align}\label{eq:condNum}
\dot{N}_1&= +2J\sqrt{N_1 N_2}\sin{\Delta\varphi} + \tfrac{N_1 U}{2}(e^{-2\ii\varphi_1}{g_{11}}^{} + \text{c.c.}),\nonumber\\
\dot{N}_2&= -2J\sqrt{N_1 N_2}\sin{\Delta\varphi} + \tfrac{N_2 U}{2}(e^{-2\ii\varphi_2}{g_{22}}^{} + \text{c.c.}),
\end{align}
which resonates with the results from Ref.~[\onlinecite{Smerzi1997}], with the additional contributions from the fluctuations. It should be noted here that $J<0$ in our convention. 

One clearly sees from \eqref{eq:flucsNum} and \eqref{eq:condNum} that the total particle number $N$ is conserved,
\begin{align}
\partial_t\sum_{\alpha} \left(N_\alpha + \delta N_\alpha\right)= \dot{N} =0\ .
\end{align}
Similarly, by employing the dynamical equations \eqref{eq:G_diag}, \eqref{eq:anomG} and \eqref{eq:G_offDiag}, the total energy
\begin{align}
E=\tfrac{1}{2}\sum_{\alpha}\, \left( E_{\alpha}^{c} + E_{\alpha}^{q}\right),
\end{align}
with the condensate energy 
\begin{align}
	\begin{split}
		E_{\alpha}^{c}=2UN_\alpha^2 + 2U F_{\alpha}N_\alpha &+ \tfrac{U}{4}\left( \ii g_{\alpha\alpha}\Phi_{\alpha}^{*}\Phi_{\alpha}^{*} + \text{c.c.}\right) \\
        &+J\left(\Phi_1^{*}\Phi_2^{} +\text{c.c.}\right),
	\end{split}
\end{align}
and the fluctuation energy
\begin{align}
	\begin{split}
		E_{\alpha}^{q}&=U F_{\alpha} \left(2N_\alpha + F_{\alpha} \right) + \tfrac{U}{2} g_{11}^{*}g_{11}^{} \\
        &+ \tfrac{U}{4}\left( \ii g_{\alpha\alpha}\Phi_{\alpha}^{*}\Phi_{\alpha}^{*} + \text{c.c.}\right) +\ii J\left( G_{12} - G_{12}^{*}\right),
	\end{split}
\end{align}
may be shown to be conserved,
\begin{align}
\ii\partial_t E=0.
\end{align}

\section*{APPENDIX B: COMPUTATION OF TRAP PARAMETERS}\label{sec:app_pots}

\subsection{Diagonalization of trap potentials} 

The trap potential employed in experiment Ref.~[\onlinecite{Albiez2005}] reads
{\small
\begin{align}
V_A({\bf r})=\tfrac{m}{2}\left[ \omega_{x}^{2}x^{2} + \omega_{y}^{2}y^{2} + \omega_{z}^{2}z^{2} \right] + \tfrac{V_0}{2}\left[ 1 + \cos{\left(\tfrac{2\pi x}{d}\right)} \right],
\end{align}
}%
with frequencies given in Ref.~[\onlinecite{Albiez2005}]. Since a Hamiltonian with this potential is separable, the eigenfunctions are 
the products of the eigenfunctions in each spatial dimension. Hence, the diagonalization of the 
noninteracting trap system reduces to three separate diagonalizations, which can be performed by applying standard library methods (e.g. Ref.~[\onlinecite{Guennebaud2010}]), yielding all eigenvalues and eigenfunctions of the trap.

\begin{figure}[h]
\centering    
\subfigure{\label{fig:wavefunction_1}\includegraphics[height=4.9cm]{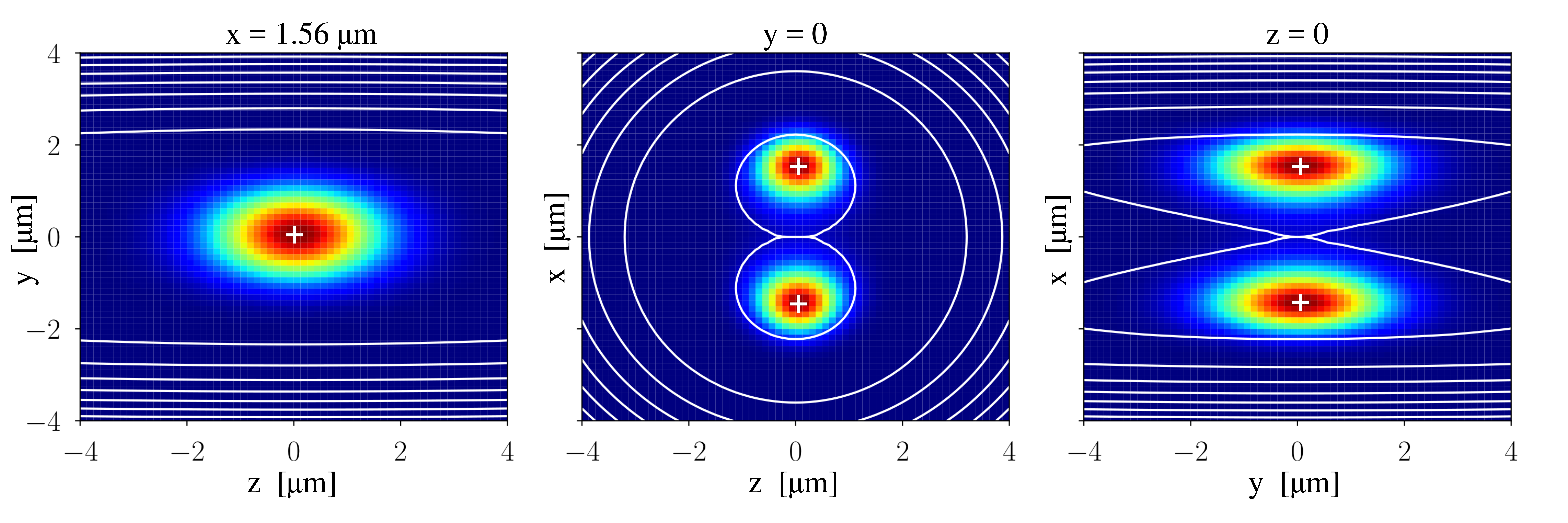}}
\vspace*{-0.2cm}
\subfigure{\label{fig:wavefunction_2}\includegraphics[height=4.9cm]{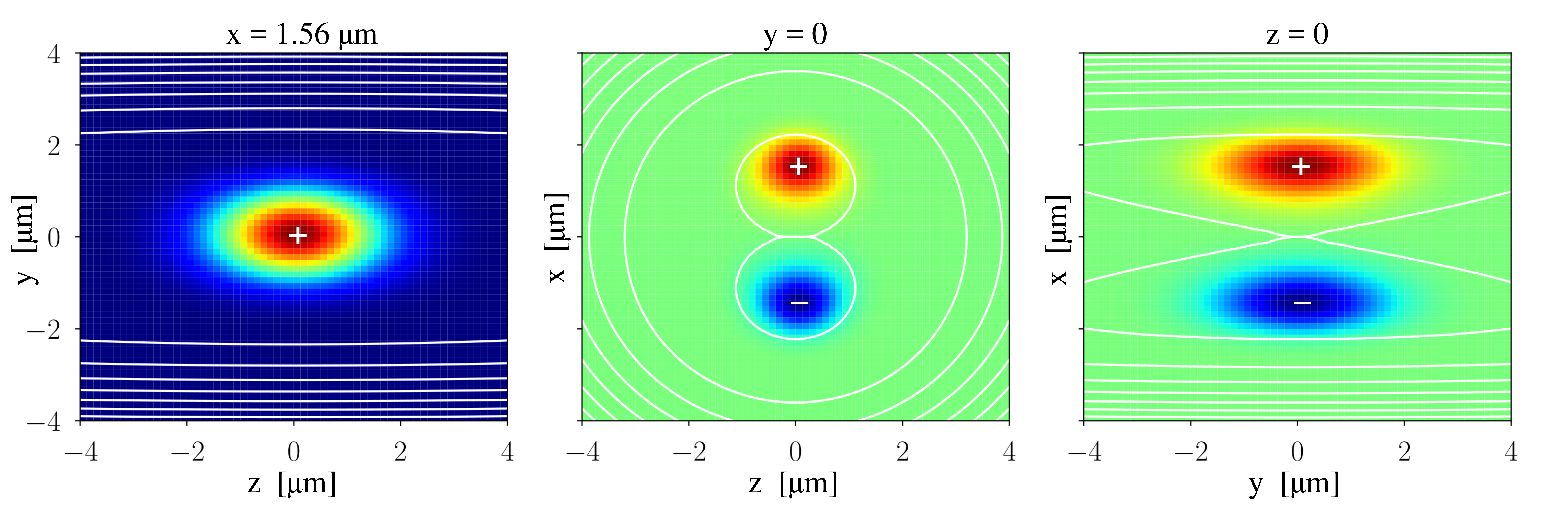}}
\vspace*{-0.2cm}
\subfigure{\label{fig:wavefunction_3}\includegraphics[height=4.9cm]{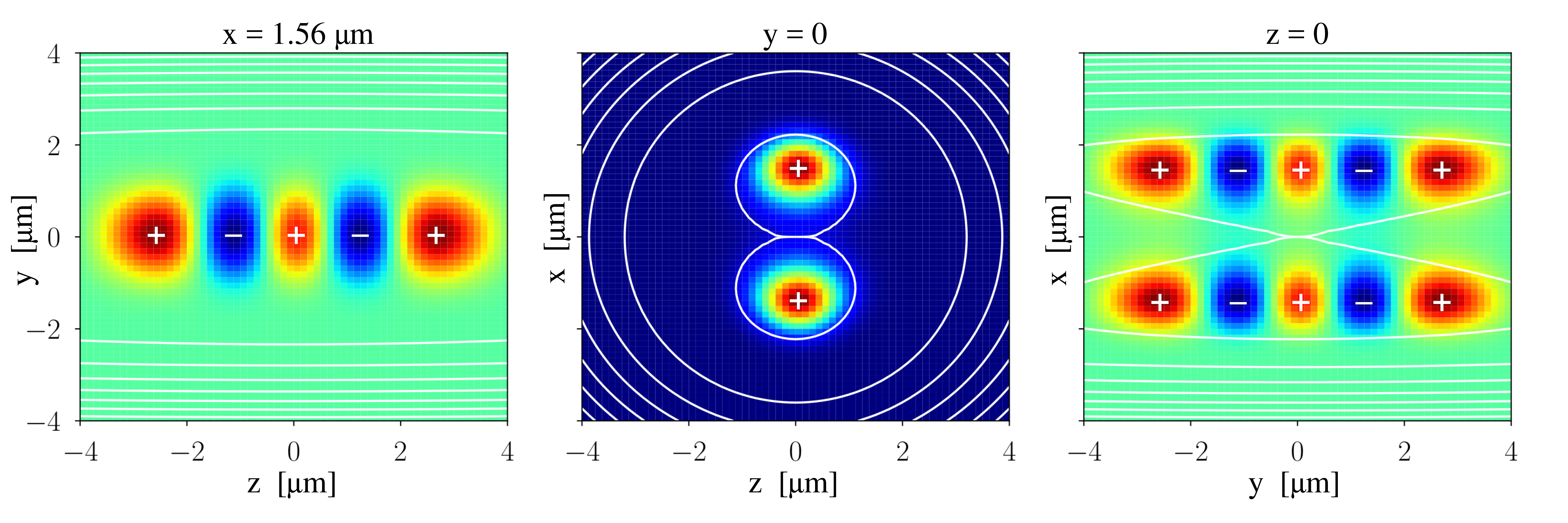}}
\vspace*{-0.3cm}
\caption{Spatial profiles of eigenfunctions of the trap potential 
Eq.~\eqref{eq:thy}, where the double-well is orianted along the $z$ axis.
Top row: the symmetric state $\varphi_+(\vect{r})$;
middle row: the antisymmetric state $\varphi_-(\vect{r})$; bottom row: 
the symmetric state corresponding to the uppermost thick level in 
Fig. \ref{fig:levels} for $a_s=0$. 
The cuts shown are along the $y-z$ plane for $x=x_{min}=1.56~\mu$m 
(position of the trap minimum), along the $z-x$ plane for $y=0$, 
and along the $x-y$ plane for $z=0$, respectively, as indicated. 
The color or gray scale describes the wave function amplitude 
(arbitrary units). $+$ ($-$) signs indicate the regions of 
positive (negative) extrema of the wave function. 
In the panels without sign change, the zero level of the wave function is 
represented by dark blue (dark gray), while in the panels with sign change, 
the zero level is represented by light green (light gray).
Some of the trap equipotential lines are shown as white 
lines, providing a guide to the eye where the minima of the 
trap potential are located. 
This solution was obtained with the Jacobi-Davidson 
algorithm for a spatial resolution of $64$ grid points in each spatial 
dimension.}
\label{fig:orbitals}
\end{figure}

The confining potential of the experiment Ref.~[\onlinecite{LeBlanc2011}] is more involved, 
{\small
\begin{align}\label{eq:thy}
V_B({\bf r})&=m'_{F}\hbar\sqrt{ \delta(\mathbf{r})^2 + \left( \frac{ \mu_{\text{B}}g_F B_{\text{RF},\perp}(\mathbf{r})}{2\hbar}\right)^2} + \tfrac{m}{2}\omega_{y}^2 y^2,
\end{align}
}%
where $\delta(\mathbf{r})=\omega_{\text{RF}} - \left|\mu_{\text{B}} g_F B_{S}(\mathbf{r})/\hbar\right|$, see Ref.~[\onlinecite{LeBlanc2011}]) for details and 
the definition of the parameters. We use the parameter values quoted there 
with $\delta = 2\pi\times (-0.4)$.
Since Eq.~\eqref{eq:thy} is not separable along the spatial axes, the  
Hamiltonian dimension is too large for direct numerical diagonalization.
In order to be as close to the actual experiment as possible, 
we expressly do not approximate Eq.~\eqref{eq:thy} by an expression 
that would be easily accessible numerically.
Therefore, one has to resort to an algorithm that can handle very large 
matrices. 
We employ the Jacobi-Davidson algorithm.\cite{Geus2002,Davidson1975}
It is an iterative subspace method that iteratively returns the first 
few eigenvalues and eigenvectors of a high-dimensional problem. 
Note that for the present analysis it is essential to include higher 
trap states. As examples of the results, the wave functions of three 
different trap eigenstates for the nonseparable potential, 
Eq.~\eqref{eq:thy}, are shown in Fig.~\ref{fig:orbitals}.

\subsection{Computation of the interaction parameters}

For the experiments (A) and (B), many of the parameters of Eq.~\eqref{eq:U} turn out to be negligible, such that, retaining only the significant parameters, the interacting part of the action can be simplified to
$
S_{\text{int}} = -\tfrac{1}{2}\left(S_{\text{loc}} + S_{12} + S_{mn} + S_{\alpha m}\right),
$
with
{\small
\begin{align}\label{eq:S_loc}
\begin{split}
S_{\text{loc}}=&\int \text{d}t\left( U \sum_{\alpha=1}^{2} \,\left|\phi_{\alpha}^{}\right|^4 + \sum_{m=3}^{M} U_m\left|\phi_{m}^{}\right|^4\right), 
\end{split}
\end{align}
\begin{align}\label{eq:S_12}
\begin{split}
S_{12}=& \int \text{d}t\; \left[\right. U'\left( {\phi_{1}^{*}\phi_{1}^{*}}{\phi_{2}^{}\phi_{2}^{}} + 2\,{\phi_{1}^{*}\phi_{1}^{}}{\phi_{2}^{*}\phi_{2}^{}} \right) \\
&+  2 J'\left({\phi_{1}^{*}\phi_{1}^{*}\phi_{1}^{}}{\phi_{2}^{}} +{\phi_{2}^{*}\phi_{2}^{*}\phi_{2}^{}}{\phi_{1}^{}}\right) + \text{c.c.} \left.\right],
\end{split}
\end{align}
\begin{align}\label{eq:S_mn}
S_{mn}= \int \text{d}t\;  \sum_{\substack{m,\,n=3\\ m\neq n}}^{M}   & \tfrac{1}{2} U'_{mn}(  {\phi_{m}^{*}\phi_{m}^{*}}{\phi_{n}^{}\phi_{n}^{}}  + 2\,{\phi_{m}^{*}\phi_{m}^{}}{\phi_{n}^{*}\phi_{n}^{}} + \text{c.c.} ) ,
\end{align}
\begin{align}\label{eq:S_am}
S_{\alpha m}=& \int \text{d}t\;\sum_{\alpha=1}^{2}\sum_{m=3}^{M} \left[\right. K_m\left( {\phi_{\alpha}^{*}\phi_{\alpha}^{*}}{\phi_{m}^{}\phi_{m}^{}} + 2\,{\phi_{\alpha}^{*}\phi_{\alpha}^{}}{\phi_{m}^{*}\phi_{m}^{}}\right) \nonumber \\
&+  2 R_{\alpha m}{\phi_{\alpha}^{*}\phi_{\alpha}^{*}\phi_{\alpha}^{}}{\phi_{m}^{}} + \text{c.c.} \left.\right].
\end{align}
}
The parameters introduced in Eqs. \eqref{eq:S_loc} -- \eqref{eq:S_am} are defined by
\begin{align}\label{eq:int_U1}
U &=\tilde{g}\int \text{d}^3 r\,\varphi_\alpha^{4}(\vect{r}), \quad \alpha = 1,\, 2 \\
U_{m} &=\tilde{g}\int \text{d}^3 r\,\varphi_m^{4}(\vect{r}),\quad m \geq 3 \\
U' &=\tilde{g}\int \text{d}^3 r\,\varphi_1^{2}(\vect{r})\varphi_2^{2}(\vect{r})
\end{align}
\newpage
\vspace*{-0.55cm}
\begin{align}\label{eq:int_U2}
J'&=\tilde{g}\int \text{d}^3 r\,\varphi_1^{3}(\vect{r})\varphi_2^{}(\vect{r})\\
U'_{mn}&=\tilde{g}\int \text{d}^3 r\,\varphi_m^{2}(\vect{r})\varphi_n^{2}(\vect{r}), \quad m,\,n \geq 3 \\
K_{m}&=\tilde{g}\int \text{d}^3 r\,\varphi_\alpha^{2}(\vect{r})\varphi_m^{2}(\vect{r}), \quad \alpha = 1,\, 2; \ m \geq 3  \\
R_{\alpha m}&=\tilde{g}\int \text{d}^3 r\,\varphi_\alpha^{3}(\vect{r})\varphi_m^{}(\vect{r}), \quad \alpha = 1,\, 2; \ m \geq 3
\end{align}

\onecolumngrid

\vspace*{0cm}

\twocolumngrid

\end{document}